\documentclass[two-columns]{nature}

\usepackage{amsmath,physics}

\usepackage{graphicx,color}
\usepackage{textcomp}

\usepackage{siunitx}

\usepackage[hidelinks]{hyperref}

\usepackage{lineno}

\usepackage{pgfplots}
\pgfplotsset{compat=1.18}
\usepackage{tikz}

\usepackage[backend=biber, bibencoding=utf8, style=nature, url=false,
    isbn=false, doi=false, maxcitenames=30, maxnames=30]{biblatex}
\author{S. Rodionov$^{1\dag}$, A. Burguete Lopez$^{1\dag}$, M. Makarenko$^{1,2}$,
Q. Wang$^{1}$, F. Getman$^{1}$, \& A. Fratalocchi$^{1*}$}

\begin{document}

\title{MOCLIP: A Foundation Model for Large-Scale Nanophotonic Inverse Design}

\maketitle

\begin{affiliations}
\item{PRIMALIGHT, Faculty of Electrical Engineering; Applied Mathematics and
    Computational Science, King Abdullah University of Science and Technology,
    Thuwal 23955-6900, Saudi Arabia
    }
\item{
    Sber AI Lab, Moscow, Russia\\
    $^\dag$ These authors contributed equally.\\
    $^*$ Corresponding author. Email: andrea.fratalocchi@kaust.edu.sa.
    }
\end{affiliations}


\begin{abstract} 
Foundation models are transforming artificial intelligence by enabling generalizable, data-efficient solutions across diverse domains and applications. However, the lack of large and diverse datasets remains a key barrier to their development in nanophotonics. This work presents MOCLIP (Metasurface Optics Contrastive Learning Pretrained), a nanophotonic foundation model that encodes metasurfaces' structural and spectral information into a shared latent space via contrastive learning, using an experimentally acquired dataset with sample density approaching the scale of ImageNet-1K. The study demonstrates MOCLIP's inverse design capabilities, including high-throughput zero-shot prediction at $2 \cdot 10^5$ samples per second,  full-wafer design of an entire 4-inch substrate in minutes, and latent space optimization reaching $95\%$ accuracy. This work further introduces an optical information storage concept that leverages MOCLIP to achieve a storage density of \num{0.1} $\text{Gbit/mm}^2$ at the resolution limit, surpassing commercial optical media by a factor of six. Together, these results position MOCLIP as a scalable, versatile platform for next-generation photonic design and data-driven applications.
\end{abstract}

\section*{Introduction}
\begin{figure}[htb]
    \centering
    \includegraphics[width=\linewidth]{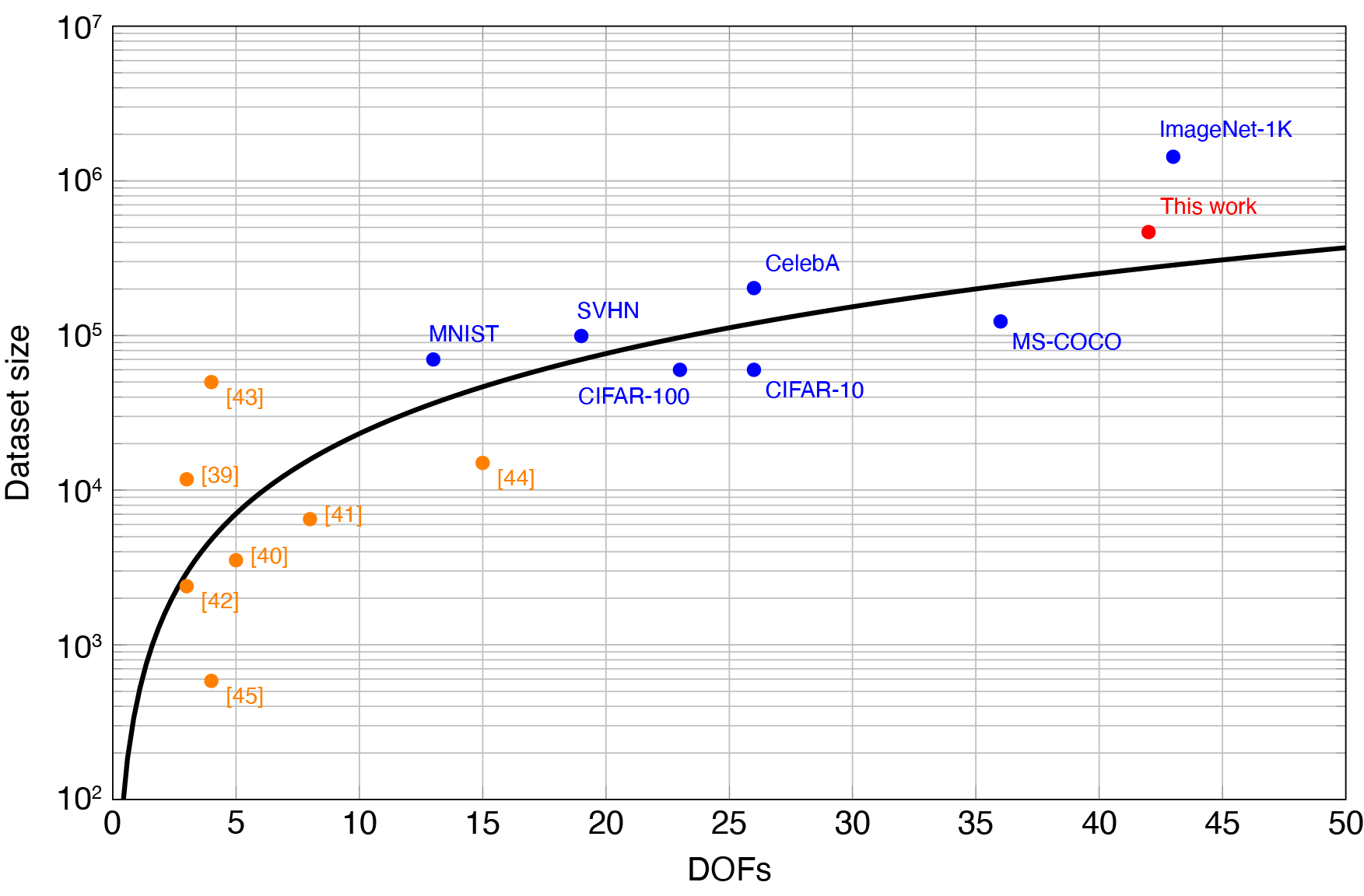}
    \caption{\textbf{State-of-the-art nanophotonic and computer vision datasets.} Dataset size versus degrees of freedom (DOFs), illustrating empirical scaling trends. Orange markers represent nanophotonic datasets, blue markers represent computer vision datasets, and this work is indicated by a red marker. Computer vision dataset sizes and DOFs are taken from~\cite{pope2021intrinsic}. The scaling law is described by the equation $N=kd^{\alpha}$, with fitted parameters $\alpha = 1.72$ and $k = 442$.}
    \label{fig:benchmark}
\end{figure}

Deep learning has rapidly permeated nanophotonic inverse design, enabling advances in metasurfaces~\cite{li2022empowering, sunae2023revisiting, wang2024transformational, seo2024deep, getman2021broadbandprima, wang2022advancing, yeung2022deepadjoint}, photonic crystals~\cite{deng2024inverse}, plasmonics~\cite{he2019plasmonic,adibnia2024deep,mahadi2024gated}, and photonic integrated circuits~\cite{maclellan2024inverse}. Beyond design, deep learning is fundamentally reshaping how optical information is processed and utilized, unlocking applications spanning Tb/s hyperspectral video understanding~\cite{makarenko2024hardware}, computational imaging~\cite{wang2025computational}, and optical sensing~\cite{alashwal2023deep}. These developments reflect a broader trend of integrating increasingly powerful artificial intelligence (AI) paradigms into photonics~\cite{wang2024transformational, ma2024optogpt, choi2024realization}. Foundation models (FMs) are at the forefront of this trend~\cite{liang2024foundation, zhou2024comprehensive, fei2022towards}. FMs are large-scale models trained in a self-supervised manner on broad data that adapt to a wide range of downstream tasks; they span architectures that include large language models (LLMs), multi-modal systems and domain-specific designs~\cite{bommasani2021opportunities, wiggins2022opportunities}. Unlike bespoke, task-specific deep learning pipelines, FMs leverage large and diverse datasets to learn universal representations that generalize across tasks and domains, enabling capabilities such as zero-shot inference and cross-domain transfer. This generalized approach is driving innovation across machine vision~\cite{awais2025foundation}, medical diagnosis~\cite{moor2023foundation, zhou2023foundation, wang2024pathology}, climate modeling~\cite{nguyen2023climax}, material chemistry~\cite{ahmad2022chemberta, batatia2023foundation}, and robotics~\cite{firoozi2023foundation}. In nanophotonics, FMs could transform inverse design by learning unified representations that generalize across devices, enabling automated design and data-driven analysis of increasingly complex nanophotonic architectures~\cite{wang2024transformational}.\\
A primary hurdle to implementing FMs in nanophotonics is their considerable training data demands~\cite{kang2024large}. Robust generalization performance requires high dataset variability, quantified by the number of degrees of freedom (DOFs) of the nanostructures~\cite{huang2020understanding,pope2021intrinsic,gong2019intrinsic}, or alternatively by the intrinsic dimension (ID) of the data manifold~\cite{gong2019intrinsic,camastra2016intrinsic}. The relationship between the dataset size $N$ required to train a deep learning model efficiently and the DOFs $d$ follows the power law $N=kd^{\alpha}$, where $k$ is a constant and $\alpha$ ranges approximately between \num{0.1} and \num{2} depending on the modality and task~\cite{sharma2022scaling, hestness2017deep, cherti2023reproducible,pope2021intrinsic}. Figure~\ref{fig:benchmark} compares state-of-the-art nanophotonic and computer vision datasets by size and DOF, revealing a substantial gap between them. Computer vision datasets, such as ImageNet‑1K, used to train large-scale FM transformers, exhibit DOFs of approximately \num{40}, yielding a density of \num{30000} samples per DOF~\cite{pope2021intrinsic}. In contrast, nanophotonic datasets typically have fewer than \num{10} DOFs and contain about \num{10} to \num{1000} samples per DOF~\cite{grbcic2024inverse,yao2023inverse,liu2018generative,roberts2021deep,wang2023_inn_metasurface,malkiel2018_plasmonic_dl,clini2023_backprop_mvall_hsis}, resulting in sample densities per DOF that fall orders of magnitude below the data densities required for FM training~\cite{fei2022towards}. A major difficulty in increasing the scale and diversity of nanophotonic datasets arises from their dependence on computationally intensive electrodynamic simulations~\cite{li2022empowering, kang2024large} based on either first principles solvers such as finite-difference time-domain (FDTD) or the finite element method (FEM), or semi-analytic techniques like rigorous coupled-wave analysis (RCWA). For example, such simulations based on FDTD with a 5~nm mesh size and a unit-cell period of 500~nm yield approximately one sample per minute, even when conducted on high-performance computing platforms~\cite{li2022inverse}. Beyond their slow data generation rate, simulations often fail to accurately capture real material properties and structural imperfections, and rarely account for response deviations from the plane-wave approximation~\cite{wang2022advancing, kang2024large, yoon2021maxim}, limiting the generalization of deep learning models trained on simulated data.\\ 
This work introduces the Metasurface Optics Contrastive Learning Pretrained (MOCLIP) framework for nanophotonic broadband amplitude inverse design. MOCLIP builds on the Contrastive Language-Image Pretrained (CLIP) foundation model~\cite{radford2021learning}, a backbone of widely adopted generative systems such as Stable Diffusion~\cite{rombach2022high}, generalizing it to the domain of nanophotonics by leveraging its ability to align multi-modal representations within a shared latent space. MOCLIP provides a unified latent space for metasurface geometries and their spectral responses, enabling both spectrum-to-geometry retrieval (inverse design) and geometry-to-spectrum retrieval (spectrum prediction)~\cite{chen2022artificial} within a single framework via latent space search or optimization.\\
MOCLIP is pretrained on broadband spectral amplitude data, whose inverse design constitutes a major open problem of this area, at the center of significant research efforts~\cite{kuznetsov2024roadmap, zeng2025performance, li2022empowering}. Addressing this issue opens a wide range of applications, including but not limited to structural color and color-printing metasurfaces~\cite{ref4}, perfect absorbers~\cite{ref5}, single-shot polarimetric imaging via dichroic metasurfaces~\cite{ref6}, display-oriented spectral filters~\cite{ref7}, compact hyperspectral imaging~\cite{makarenko2024hardware, ref9}, humidity-responsive structural color pixels~\cite{ref10}, biosensors~\cite{ref11, ref12}, and temperature sensors~\cite{ref13}.\\
MOCLIP undergoes training entirely on experimental data, without relying on computer simulations, using a dataset generation strategy that combines dense wafer-scale fabrication with automated hyperspectral measurements. This pipeline enables fabrication of around \num{40} samples per minute and optical characterization at a rate of \num{380} samples per minute. The model uses more than \num{11000} samples per DOF, approaching the scale of ImageNet-1K and exceeding the sample density of the majority of state-of-the-art computer vision datasets. This work showcases different applications of MOCLIP, including two complementary inverse design modalities: large-scale zero-shot prediction at a rate of $2 \cdot 10^5$ samples per second and latent space optimization for high-fidelity inverse design with over $95\%$ accuracy, as well as a proof-of-concept optical information storage technology with information density up to \num{0.1} $\text{Gbit/mm}^2$, surpassing the information density of commercial optical storage media by a factor of six~\cite{blu-raydiscassociationWhite2010}.

\section*{Results}

\begin{figure}[htb] \centering
	    \includegraphics[width=\linewidth]{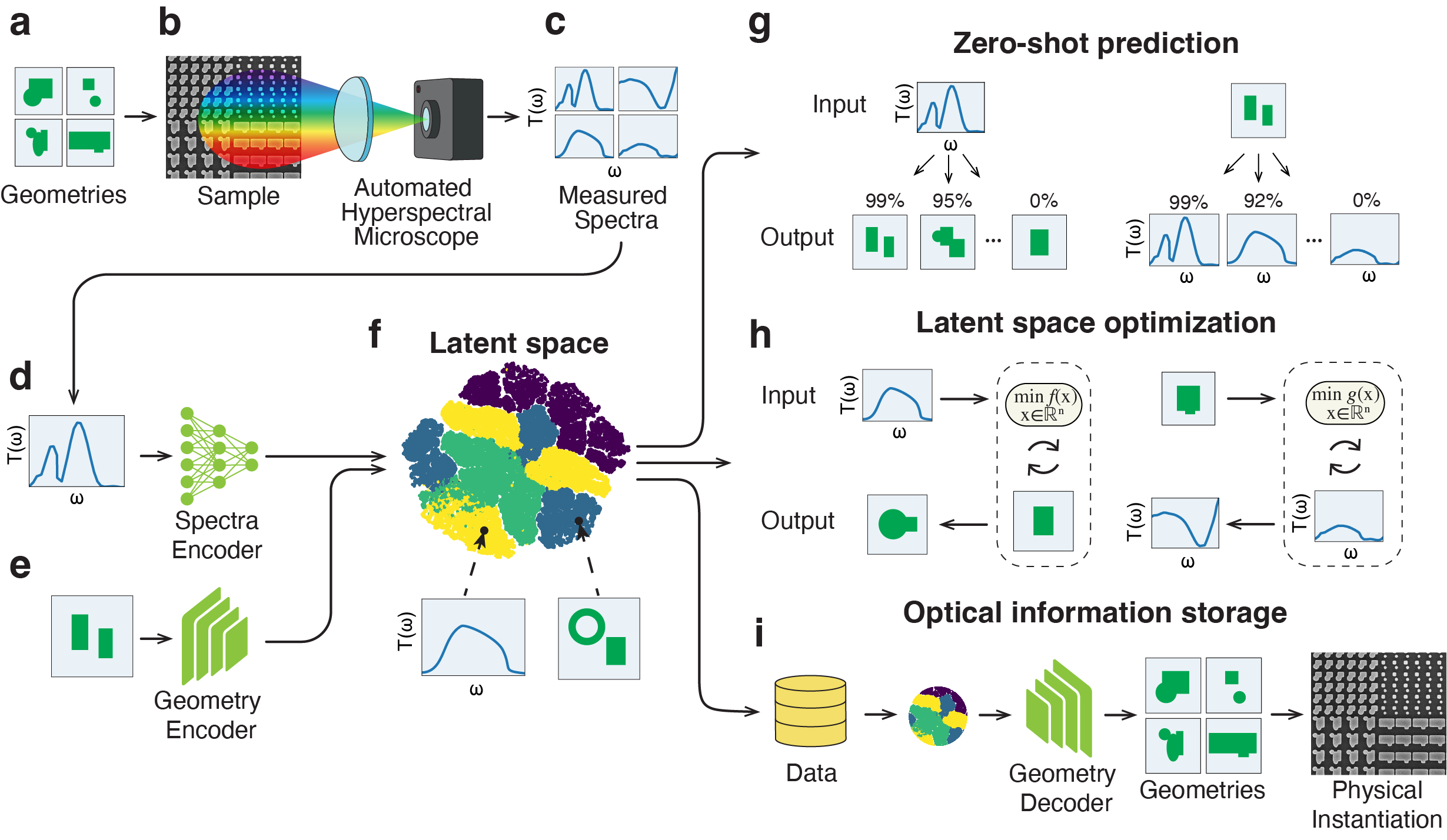}
	    \caption{\textbf{MOCLIP concept.} \textbf{a} Dataset of randomly generated free-form metasurface geometries. \textbf{b} Experimental realization and characterization of the designs using an automated hyperspectral microscope. \textbf{c} Dataset of measured spectra corresponding to the free-form metasurface designs. \textbf{d} Encoding of experimental metasurface spectral responses. \textbf{e} Encoding of metasurface geometries. \textbf{f} Shared latent space for spectral and geometric information. \textbf{g} Zero-shot prediction for inverse design (left) and spectra prediction (right). \textbf{h} Latent space optimization for inverse design (left) and spectra prediction (right). \textbf{i} Optical information storage via physical implementation of latent space information.}
        \label{fig:concept}
\end{figure}

Figure~\ref{fig:concept} summarizes MOCLIP's dataset generation, high-level architecture, and applications. Dataset generation starts by constructing a library of free-form metasurface geometries assembled from shape primitives: ellipses, rectangles, and rings (Fig.~\ref{fig:concept}a). Subsequent fabrication transfers these designs into silicon-on-glass metasurfaces, accommodating thousands of distinct geometries on a single substrate. A custom automated hyperspectral transmission microscope measures each metasurface (Fig.~\ref{fig:concept}b). Figure~\ref{fig:concept}c illustrates the measurement results, comprising spectral responses for each metasurface geometry. The union of all geometries and their corresponding spectral responses forms MOCLIP's training dataset.\\
Following the CLIP FM architecture, the framework trains two distinct encoders on input data pairs comprising experimental spectral response curves and binary geometry images (Fig.~\ref{fig:concept}d-e). MOCLIP uses a multilayer perceptron (MLP) for the spectral response curves and a convolutional neural network (CNN) for the geometry images. The model projects the encoders' outputs into a shared low-dimensional latent space (Fig.~\ref{fig:concept}f). MOCLIP's training aligns the latent vectors of matching geometry (A) and spectrum (B) pairs by maximizing the cosine similarity $S_C$ between their latent vectors, while simultaneously minimizing the similarity between all non-matching (off-diagonal) pairs:
\begin{equation}
\label{eq:cosine_similarity}
 S_C(\vb{A},\vb{B})= \frac{\vb{A}\vdot\vb{B}}{\norm{\vb{A}}\norm{\vb{B}}}
\end{equation}
Once trained, MOCLIP generalizes to a broad range of applications in broadband amplitude inverse design, including zero-shot prediction, latent space optimization, and automated optical information encoding and storage, all with little to no additional fine-tuning (Fig.~\ref{fig:concept}g-i).

\subsection*{Dataset generation}
\begin{figure}[htb] \centering
	    \includegraphics[width=\linewidth]{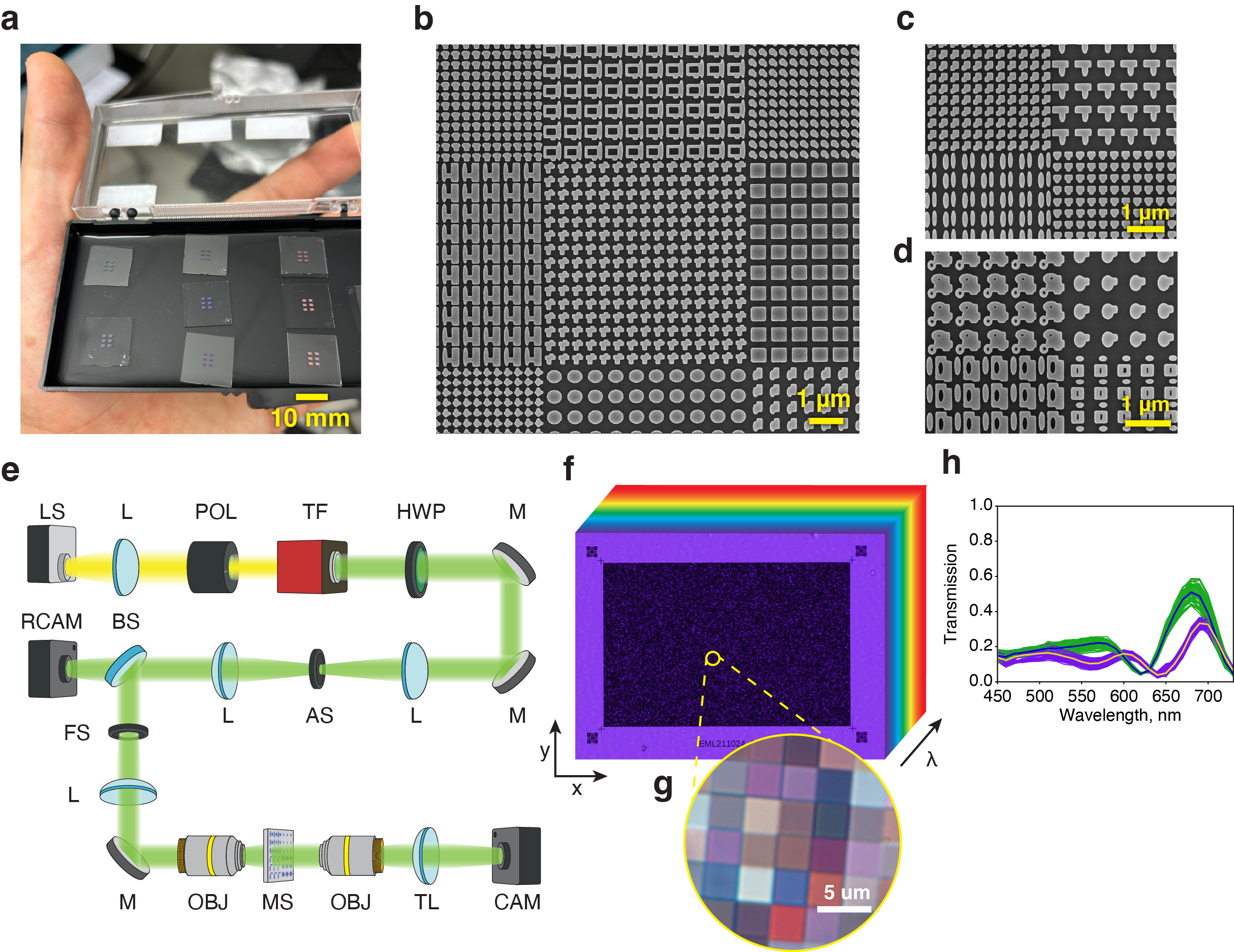}
	\caption{\textbf{Dataset generation.}
    \textbf{a} A set of fabricated samples, each carrying six metasurface arrays of amorphous silicon on a fused-silica substrate, totaling \num{136 530} dataset samples per substrate.
    \textbf{b-d} SEM images of the fabricated metasurface arrays.
    \textbf{e} Hyperspectral transmission microscopy setup.
    \textbf{f} Spatial and spectral data hypercube of a metasurface array under fixed polarization conditions, shown in false colors for visualization.
    \textbf{g} Expanded view of the metasurface array under broadband illumination.
    \textbf{h} Example spectra extracted from the pixels of a single metasurface pad for x- and y-polarized illumination (green and violet sets of curves) and average pixel responses (black and yellow curves).
    }   
        \label{fig:dataset}
\end{figure}

Dataset generation starts with a stochastic algorithm that combines shape primitives to generate free-form design geometries (see Methods). The generation algorithm enforces dimensional limits and verifies element placement to ensure compliance with fabrication process tolerances. Metasurface fabrication involves electron beam lithography patterning of amorphous silicon layers on fused silica substrates (see Methods). Figure~\ref{fig:dataset}a shows all the fabricated samples used for measurements and dataset collection. There are 8 substrates in the picture, each holding 6 metasurface arrays visible as individual chips. Every metasurface array comprises \num{22755} individual metasurfaces of size 6~$\mu$m $\times$ 6~$\mu$m, totaling \num{136530} metasurfaces per substrate. In total, we fabricated and measured \num{136530} $\cdot \ 8 =$ \num{1092240} metasurfaces, half of which are unique and the other half exact replicas (see Supplementary Note 1). The replicated dataset accounts for fabrication artifacts, contamination, and crosstalk from neighboring structures. Figures~\ref{fig:dataset}b-d present scanning electron microscope (SEM) images of representative fabricated metasurface arrays of different geometries. Quality assessment involves fabricating a companion $10\times 10$ metasurface array containing randomly sampled entries for each of the 24 arrays (see Supplementary Note 3). SEM characterization of these companion arrays reveals a mean size difference of \SI{4}{nm} and a standard deviation of \SI{10}{nm} between the fabricated structures and their designs across all dataset metasurfaces.\\ 
Figure~\ref{fig:dataset}e illustrates the hyperspectral transmission microscope setup (see Supplementary Note 2) for characterizing the metasurfaces' optical responses. Light from a broadband source (LS) passes through a collimating lens (L), a polarizer (POL), and a tunable filter (TF), forming a monochromatic beam of tunable wavelength and controlled linear polarization. A half-wave plate (HWP) controls the polarization of the filter output, after which a series of mirrors (M) and lenses direct the beam through the microscope's aperture and field stops (AS, FS), focusing it onto the back focal plane of an objective (OBJ) to illuminate the sample with K\"ohler illumination. A second objective and a tube lens (TL) collect the transmitted light and image the sample onto a camera (CAM). Simultaneously, a beam splitter (BS) diverts a portion of the beam to a reference camera (RCAM), which monitors and compensates for intensity fluctuations.\\
For every array, the setup records two stacks of narrow-band images by driving the TF from \SIrange{450}{730}{nm} in \SI{10}{nm} increments at its minimum bandwidth setting (\SI{10}{nm}) for $x$ and $y$ polarizations, generating two hypercubes per array. Supplementary Notes 2 and 3 provide details on the setup components and measurement error analysis. Figure~\ref{fig:dataset}f shows a representative hyperspectral datacube acquired from a single metasurface array under fixed polarization conditions. False colors enhance visualization, with the dark rectangular region at the center corresponding to the spatial footprint of the array. Figure~\ref{fig:dataset}g provides an expanded view of the metasurface array under broadband illumination, showing the spectral response of each metasurface as distinct colors captured by a conventional red, green, and blue (RGB) camera. Figure~\ref{fig:dataset}h depicts the response of a typical metasurface computed from the hypercube's pixel values. The green and violet lines represent readings obtained from all pixels covering a single metasurface in the $x$ and $y$ polarization hypercubes, while the black and yellow lines show the averaged pixel responses used as dataset entries. Cross-validation of every metasurface spectrum with its corresponding duplicate in the replicated dataset ensures quality by discarding entries with a mean-squared error (MSE) exceeding $10^{-3}$ value, resulting in a dataset of \num{466 537} unique metasurface geometry-spectrum pairs.

\subsection*{MOCLIP training}
\begin{figure}[htb] 
        \centering
	    \includegraphics[width=\linewidth]
        {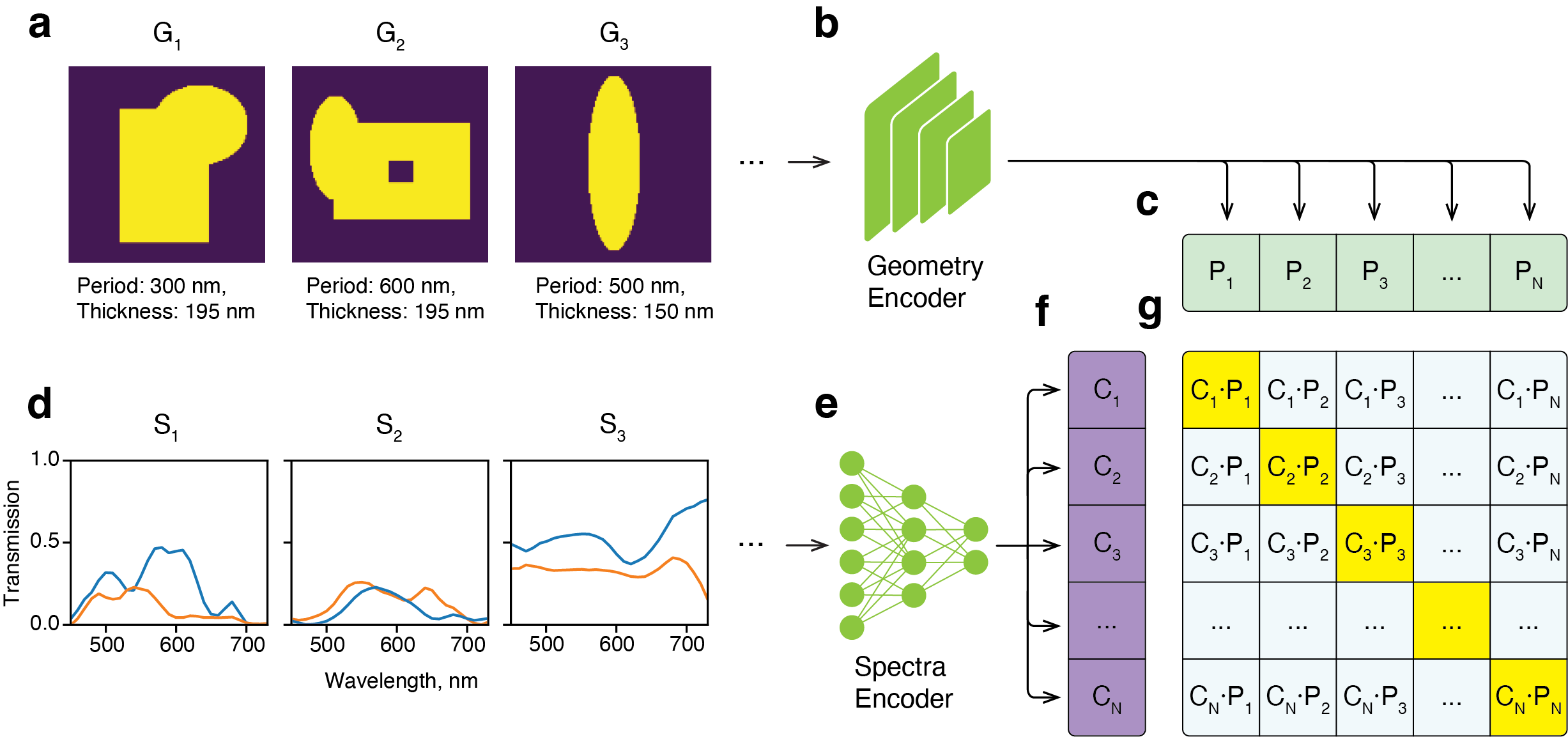}
	    \caption{\textbf{MOCLIP training.} \textbf{a} Representation of geometrical metasurface information as $128\times 128$ binary images with period and thickness parameters. \textbf{b} Geometry encoder, comprising a CNN architecture. \textbf{c} 64-dimensional latent vectors produced by the geometry encoder. \textbf{d} Spectral information represented by two 29-dimensional vectors for x- and y-polarizations (blue and orange curves, respectively). \textbf{e} Spectra encoder based on an MLP architecture. \textbf{f} 64-dimensional latent vectors produced by the spectra encoder. \textbf{g} Similarity matrix containing pairwise dot products between latent vectors from both modalities. Yellow entries correspond to matching geometry–spectra pairs trained to have similarity values close to 1, with the rest indicating non-matching pairs trained to approach zero.
        }   
        \label{fig:training}
\end{figure}

Figure~\ref{fig:training} illustrates MOCLIP's training process. The pipeline follows a bimodal contrastive learning approach, simultaneously training two encoders to project their outputs into a shared latent space. Figures~\ref{fig:training}a-c show the geometry encoding process. Binary metasurface geometry images (Fig.~\ref{fig:training}a) are processed by the geometry encoder (Fig.~\ref{fig:training}b), which uses a CNN based on the ResNet architecture~\cite{heDeep2015} to generate a latent vector $P_n$ for each geometry $G_n$, where $n$ is a sample index (Fig.~\ref{fig:training}c). Each vector $P_n$ is $M$-dimensional, where $M$ is a hyperparameter determined through systematic optimization to identify the optimal dimensionality of MOCLIP's latent space, yielding $M = 64$.\\
Figures~\ref{fig:training}d-f show the spectral information encoding process. The encoding process begins by concatenating the two polarization responses for each metasurface into a single vector $S_n$ (Fig.~\ref{fig:training}d). 
These vectors are fed into the spectra encoder (Fig.~\ref{fig:training}e), which uses an MLP architecture to produce the corresponding latent vectors $C_n$ (Fig.~\ref{fig:training}f). Supplementary Note 4 provides additional information on the MOCLIP architecture. \\
Figure~\ref{fig:training}g shows the similarity matrix obtained by computing the pairwise cosine similarities $S_C$ between the normalized latent vectors from the geometry and spectra input modalities. The cross-entropy loss~\cite{bishopPattern2016} between the similarity matrix and an identity matrix enforces contrastive learning by promoting the alignment of matching pairs while penalizing mismatches. The Methods section provides additional details on the training process. 

\subsection*{Zero-shot prediction}
\begin{figure}[htb] \centering
	    \includegraphics[width=\linewidth]
        {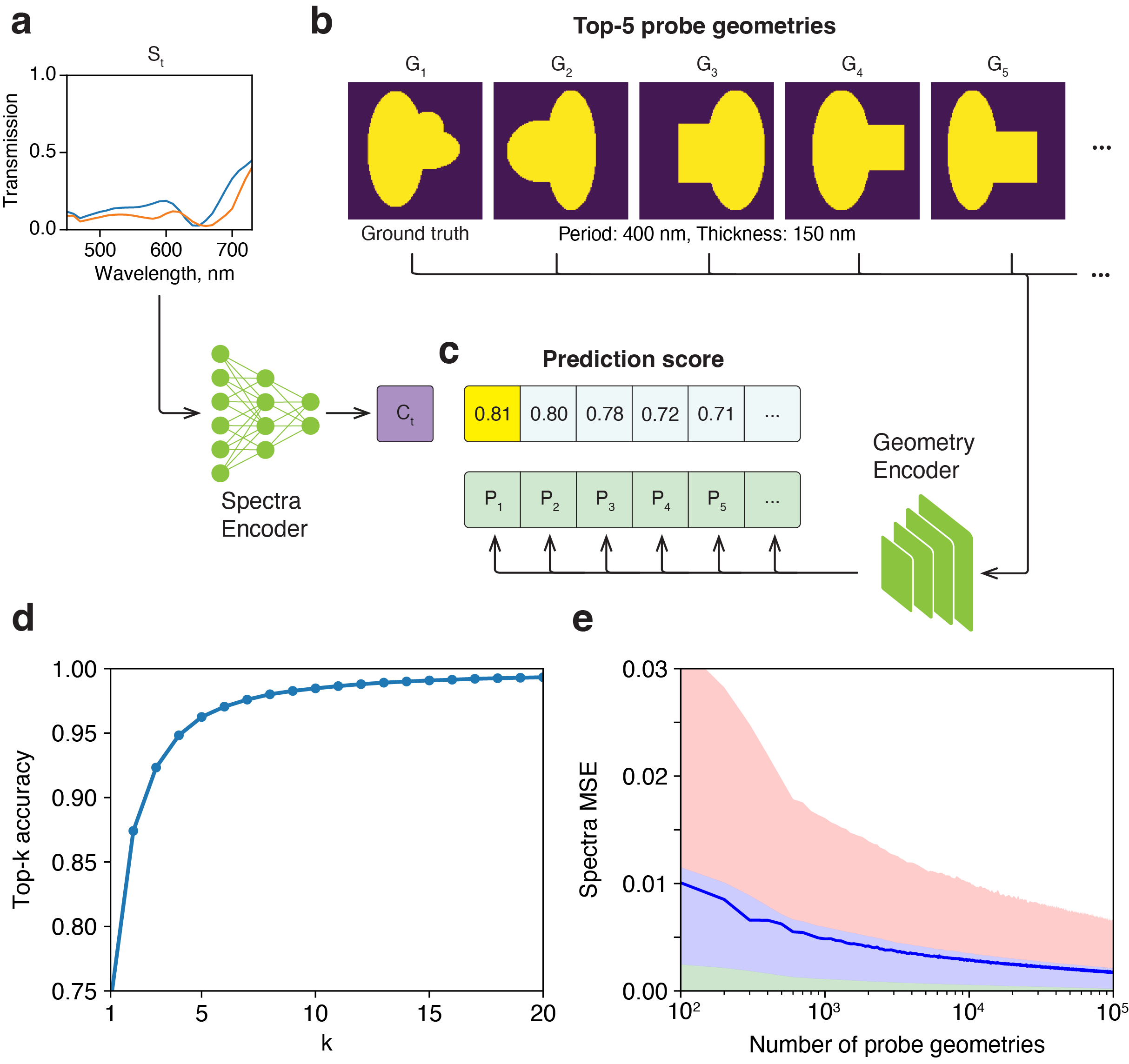}
	    \caption{\textbf{Zero-shot prediction.}
\textbf{a} Target spectrum encoding into a latent vector, where the x- and y-polarizations are represented by blue and orange curves, respectively. \textbf{b} Encoding of probe geometries, including the ground truth one, into latent vectors. \textbf{c} Similarity score vector between the target spectrum and the candidate geometry latent vectors. The yellow cell indicates the best matching score for the predicted geometry. \textbf{d} Top-k accuracy for the test dataset. \textbf{e} Statistical distribution of the MSE between the target spectrum and measured spectra of the predicted geometries as a function of the number of probe geometries. Shaded bands indicate quantile regions---green for $<$Q25, blue for Q25--Q75 and red for Q75--Q95; the solid blue line indicates the mean.}
     \label{fig:zero-shot}
\end{figure}

In deep learning, zero-shot prediction refers to a model's ability to make meaningful predictions for inputs or tasks not explicitly represented during training. This capability arises from leveraging a shared latent space that preserves semantic relationships. For example, a CLIP model trained on photographs can accurately classify stylized drawings of the same subjects~\cite{girdhar2023imagebind,radford2021learning}. This capability is a defining feature of FMs, enabling reduced annotation costs and turnaround times, sustained functionality in open-world settings where new concepts can emerge, and applicability to long-tail use cases with scarce examples~\cite{liu2023remoteclip}. MOCLIP delivers zero-shot nanophotonic inverse design capabilities, enabling rapid identification of candidate metasurface geometries that can match a previously unseen target spectral response.\\
Figures~\ref{fig:zero-shot}a-c illustrate MOCLIP's zero-shot prediction process. The process starts by encoding the target spectrum $S_t$ with the spectra encoder to produce its latent vector $C_t$ (Fig.~\ref{fig:zero-shot}a). Next, MOCLIP generates a set of probe geometries $G_n$ (Fig.~\ref{fig:zero-shot}b) and encodes each through the geometry encoder to obtain their corresponding latent vectors $P_n$. Probe geometries can be any collection of candidates, such as randomly generated free-form shapes or a pre-existing library. By computing the cosine similarity between the spectrum's latent vector and each of the probe geometries' latent vectors, MOCLIP generates a similarity score for each probe geometry, indicating how closely it matches the target spectrum (Fig.~\ref{fig:zero-shot}c). MOCLIP then selects the highest-scoring geometry (Fig.~\ref{fig:zero-shot}c, yellow cell) as its prediction.\\
We evaluate MOCLIP's zero-shot performance by calculating the Top-$k$ retrieval accuracy, a standard metric for evaluating retrieval performance in FMs~\cite{petersen2022differentiable}. This metric quantifies the proportion of queries in which a model retrieves the correct match within the first $k$ candidates of a ranked retrieval list. Figure~\ref{fig:zero-shot}d reports MOCLIP's Top-$k$ retrieval accuracy for $k$ ranging from 1 to 20 evaluated on a test set of \num{46653} samples (see Methods). For each target spectrum in the test set, the model retrieves the top $k$ most similar geometries from the test set, based on the cosine similarity of the latent vectors. MOCLIP achieves Top-1, Top-5, and Top-10 retrieval accuracies of nearly 75\%, 95\%, and 98\%, respectively. In an analogous text-to-image multi-modal retrieval task, the original CLIP model achieves Top‑1, Top‑5, and Top‑10 accuracies of 68.7\%, 90.6\%, and 95.2\% on the Flickr30k dataset, and 37.8\%, 62.4\%, and 72.2\% on the MS-COCO 5k test set, respectively~\cite{radford2021learning}. A similar evaluation in the materials science domain (predicting crystalline structure from the density of states) reports Top‑1, Top‑5, and Top‑10 retrieval accuracies of approximately 40\%, 75\%, and 85\%, respectively~\cite{moro2023multimodal}.\\
MOCLIP's zero-shot prediction offers substantially higher throughput than conventional inverse design approaches. Starting from a prepared probe set of candidate geometries generated using the same algorithm as during dataset creation, MOCLIP encodes each candidate geometry into a latent vector using the geometry encoder (Fig.~\ref{fig:zero-shot}b). This process yields a probe database comprising metasurface geometries and their latent vectors at a throughput exceeding \num{7000} designs per second for database construction, after which the database supports inverse design inference at a rate exceeding $10^9$ probe geometries per second. A key advantage of this approach is that the probe geometries can be customized and filtered during the initial database generation stage, allowing direct incorporation of fabrication tolerances or application-specific constraints without any modifications to the model architecture or retraining.\\
We evaluated the model's zero-shot prediction accuracy by comparing  experimentally acquired spectra of MOCLIP's predicted geometries against target spectra. We conducted the error analysis on the test dataset and visualized in Fig.~\ref{fig:zero-shot}e the statistical distribution of MSE between \num{10000} randomly sampled target spectra and their MOCLIP-predicted counterparts using varying probe database sizes. The shaded regions indicate quantile bands: below the 25th percentile (green), between the 25th and 75th percentiles (blue), and between the 75th and 95th percentiles (red). When the probe database size exceeds $10^4$ samples, 95\% of predictions achieve an MSE $< 10^{-2}$, establishing a 10\% transmission deviation as an upper bound for reliable inverse design~\cite{chen2022artificial}. At this database size and accuracy level, MOCLIP achieves an inverse design throughput of one high-fidelity solution every 5~µs, representing a $10^2\times$ speedup over convolutional generative adversarial network (cGAN) models~\cite{liu2018generative} and a $10^6\times$ speedup over state-of-the-art diffusion-based inverse design models~\cite{zhu2024rapid}. As a practical demonstration, MOCLIP completes the inverse design of an entire 4-inch wafer filled with 5~µm square metasurfaces in 20 minutes, a task that would typically require weeks or months at this metasurface density and resolution using state-of-the-art approaches.\\  
MOCLIP's symmetric treatment of metasurface geometries and their spectra enables zero-shot prediction of spectral responses from target geometries, effectively framing spectrum prediction as a cross-modal retrieval task. Using a database of experimentally measured spectral responses, MOCLIP replaces time-consuming electrodynamic simulations with fast zero-shot predictions, achieving speedups of up to $10^7$ (see Supplementary Note 5), while drawing on experimentally validated data. This approach achieves an accuracy comparable to that of MOCLIP's inverse design task. Supplementary Note 5 details the zero-shot spectrum prediction pipeline and reports the corresponding performance metrics. 

\subsection*{Library search}
\begin{figure}[htb] \centering
\includegraphics[width=\linewidth,height=0.78\textheight,keepaspectratio]
        {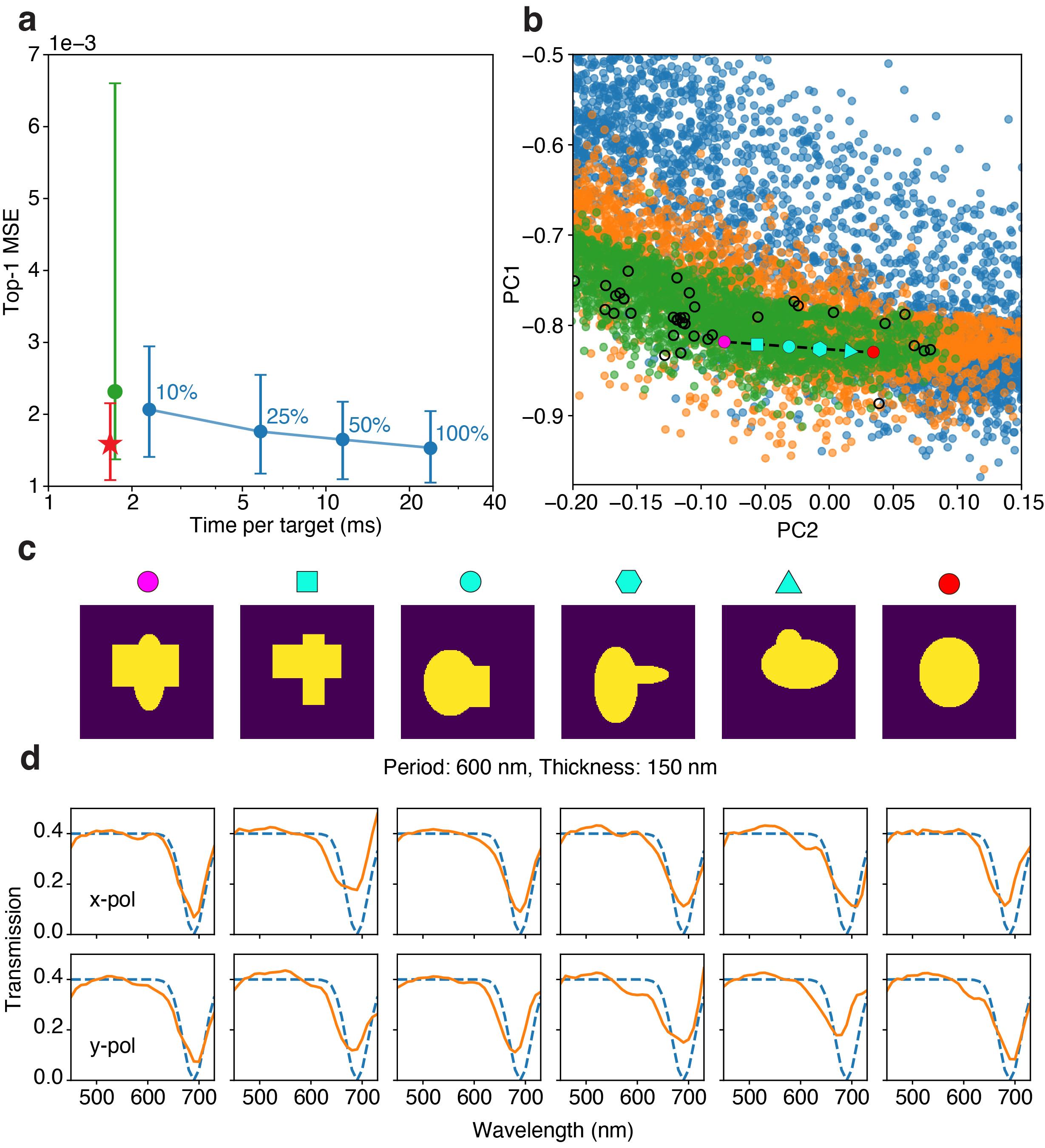}
	    \caption{\textbf{Library search.}
\textbf{a} Speed benchmark comparing exact MSE search (blue markers), spectra similarity search (green marker), and MOCLIP latent-vector search (red star). \textbf{b} Two-dimensional visualization of the geometry latent space showing the top-50 library geometries (black circles) and their nearest-neighbor variations (blue dots for low similarity, orange dots for moderate similarity and green dots for high similarity). Magenta and red markers indicate the start and end anchors for interpolation, respectively. \textbf{c} Start and end anchor geometries selected from the top-50 library candidates, together with the interpolated geometry variants. \textbf{d} Experimental spectra for the anchor geometries (first and last columns) and simulated spectra for the interpolated geometries (solid lines), compared against the target spectrum (dashed lines).}
     \label{fig:library-search}
\end{figure}

One of the most straightforward and widely used approaches for metasurface inverse design is library search~\cite{li2022inverse, li2022empowering}. In this framework, a generated database of metasurface geometries and their corresponding optical responses serves as a search space for the inverse design. Given a target optical response, we obtain the inverse-design solution by selecting the library entry whose response most closely matches the target according to a chosen similarity metric, such as MSE. Although this approach is conceptually simple and can provide highly accurate matches within the sampled database, it remains fundamentally limited by its discrete nature: it does not naturally support reliable interpolation in high-dimensional parameter spaces, and retrieving the best match directly in the original spectral space can be inefficient for large libraries~\cite{choi2024realization, li2022empowering}. MOCLIP’s zero-shot prediction capability addresses these limitations by enabling faster retrieval and continuous interpolation between neighboring geometries in the learned latent space.\\
To compare conventional library-based search with MOCLIP-based retrieval, we first generated a set of custom polarization-resolved transmission spectra corresponding to ideal broadband color filters, neutral-density filters, and polarizers (see Supplementary Note 7). In addition, to ensure a fair comparison, we precomputed latent vectors for all geometries in the database following the procedure described in the zero-shot prediction section. For each custom target spectrum, we then searched the database for the closest match by comparing the target either directly with library spectra in the original spectral space or with library geometries in the learned latent space. Specifically, we considered three retrieval strategies: MSE-based search in the original spectral space, Top-1 cosine-similarity search in the latent space, and Top-1 cosine-similarity search in the original spectral space. We included the last strategy to isolate the effect of the similarity metric from that of the representation space, thereby enabling a direct comparison between MSE- and cosine-similarity-based retrieval. For each method, we recorded the inference time per target and the distribution of reconstruction errors across all target spectra, summarized by the median MSE and the 25th and 75th percentiles.\\
Figure~\ref{fig:library-search}a shows the library search speed benchmark for the three retrieval strategies described above. Each marker indicates the median MSE across all target spectra, while the error bars denote the 25th and 75th percentiles. The blue markers correspond to conventional MSE-based library search evaluated on progressively larger fractions of the full database (10\%, 25\%, 50\%, and 100\%). When applied to the full database, conventional MSE-based retrieval requires 23 ms per target, with the MSE range between the 25th and 75th percentiles spanning from $1.05 \cdot 10^{-3}$ to $2.05 \cdot 10^{-3}$. The red star denotes MOCLIP-based retrieval in the latent space, where cosine similarity is computed between the latent vector of the target spectrum and the latent vectors of all geometries in the database. Relative to the full-database MSE search, it achieves more than a 10$\times$ reduction in inference time, reaching 1.67 ms per target, while maintaining a comparable accuracy range of $1.09 \cdot 10^{-3}$ to $2.15 \cdot 10^{-3}$. In contrast, Top-1 cosine-similarity search in the original spectral space (green marker) yields a similar inference time of 1.8 ms per target, which is within measurement uncertainty of the MOCLIP runtime, but substantially lower retrieval accuracy, with the MSE range increasing to $1.4 \cdot 10^{-3}$ - $6.6 \cdot 10^{-3}$. These results show that MOCLIP-based retrieval preserves the accuracy of conventional MSE-based library search while reducing the computational cost by more than an order of magnitude, highlighting the advantage of performing retrieval in the learned latent space rather than directly in the original spectral space.\\
Supplementary Note 9 discusses scaling times of MOCLIP latent space compared to direct library search with and without nearest neighbor acceleration. For databases of sizes $10^7$, $10^8$, and $10^9$, latent space retrieval is 6-14 times faster while preserving full accuracy in the results. Beyond runtime considerations, a key advantage of MOCLIP is the ability to identify accurate inverse-design solutions for targets that lie outside the sampled library, which conventional library search cannot address by construction.\\
Figures~\ref{fig:library-search}b–d further show that MOCLIP supports continuous interpolation between neighboring metasurface geometries in the shared latent space. To illustrate this behavior, we randomly selected one target spectrum (Fig.~\ref{fig:library-search}d, blue dashed line). We analyzed the local arrangement of its nearest library matches together with the interpolated geometries between them in the latent space. To this end, we applied principal component analysis (PCA) to the full set of geometry latent vectors. We projected them onto a two-dimensional space (PC1 and PC2) to visualize the latent-space structure (Fig.~\ref{fig:library-search}b). We then identified the Top-50 library matches for the selected target spectrum and located their corresponding geometry latent vectors within this projection~(Fig.~\ref{fig:library-search}b, black rings). As the library geometries occupy only discrete locations in the latent space, we filled the surrounding region by generating \num{10000} geometries through small random perturbations of the Top-50 geometry candidates and coloring them according to their cosine similarity to the target spectrum: green denotes similarity comparable to the library matches, orange denotes lower but still acceptable similarity, and blue denotes substantially lower similarity corresponding to poor candidates~(Fig.~\ref{fig:library-search}b). The visualization shows that cosine similarity defines a local trust region in the latent space (Fig.~\ref{fig:library-search}b, green area), where interpolation identifies suitable candidates for inverse design, including potentially better solutions than the original library matches, while effectively excluding poor candidates. For demonstration purposes, we selected two library points in the 2D projection (Fig.~\ref{fig:library-search}b, start and end anchors), connected them by a black dashed line, and extracted four geometries located along this line (Fig.~\ref{fig:library-search}b, cyan markers), which are shown in Fig.~\ref{fig:library-search}c. We additionally compared their spectra with the target spectrum in Fig.~\ref{fig:library-search}d, where we computed spectra of the interpolated geometries using FDTD. Together, these results show that MOCLIP's latent space supports physically meaningful interpolation between neighboring library geometries, extending the conventional discrete library search into a continuous local design space.

\subsection*{Design by latent space optimization}
\begin{figure}[htb] \centering
	    \includegraphics[width=\linewidth]
        {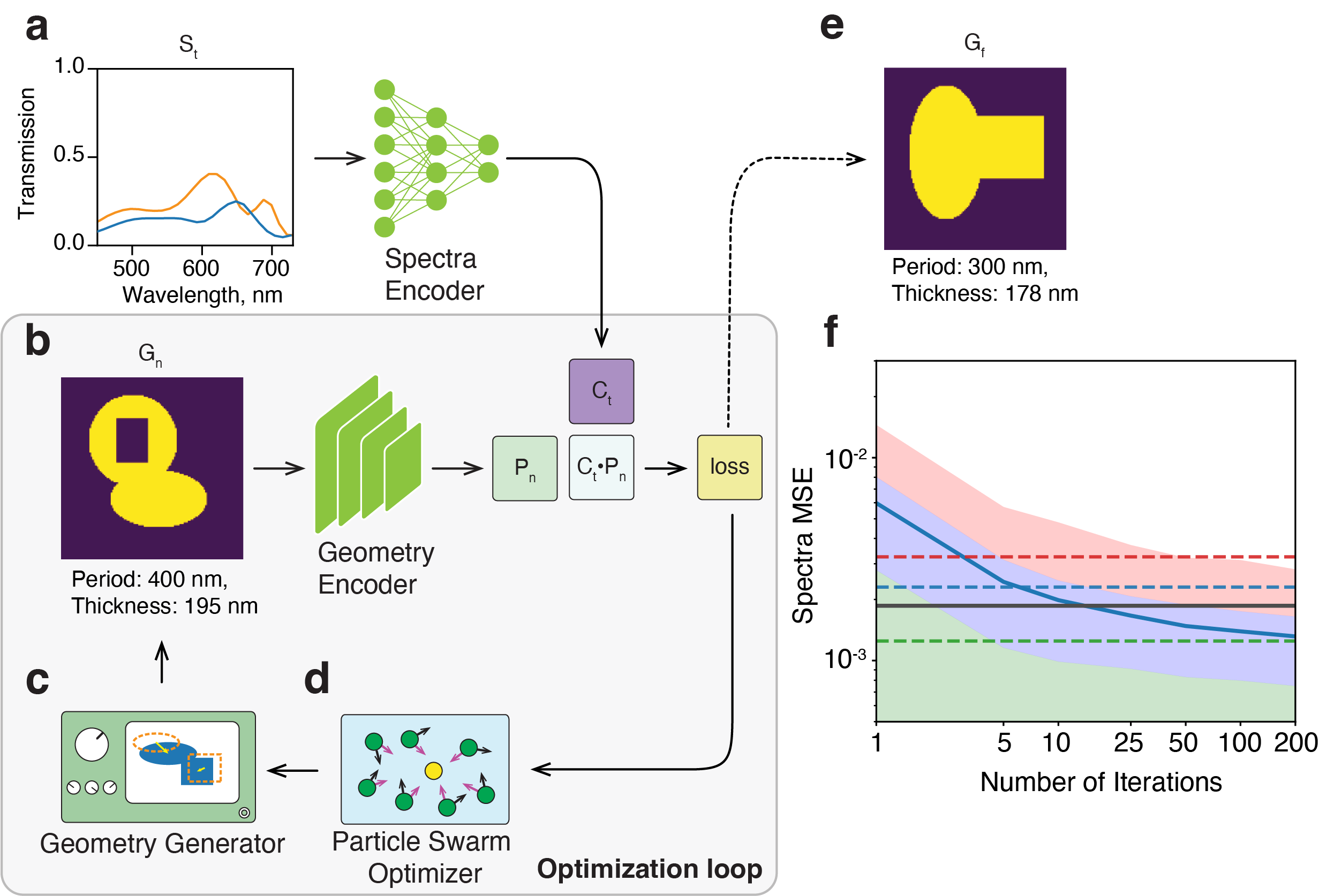}
	    \caption{\textbf{Latent space optimization.} \textbf{a} Target spectrum encoding into a latent-space vector, where the x- and y-polarizations are represented by blue and orange curves, respectively. \textbf{b} Intermediate geometry prediction encoding into a latent-space vector. \textbf{c} Geometry generator with trainable parameters. \textbf{d} Optimizer for the geometry generator. \textbf{e} Optimized geometry prediction. \textbf{f} Optimization accuracy quantified by the statistical distribution of the MSE between the target and predicted spectra as a function of the number of iterations. Shaded bands indicate quantile regions---green for $<$Q25, blue for Q25--Q75 and red for Q75--Q95; the solid blue line indicates the mean; dashed curves indicate quantile lines for the library-based search approach on the whole dataset---green for Q25, blue for Q75 and red for Q95; the black solid curve indicates the mean value for the library search.}
        \label{fig:optimization}
\end{figure}

\begin{figure}[htb] \centering
	    \includegraphics[width=\linewidth]
        {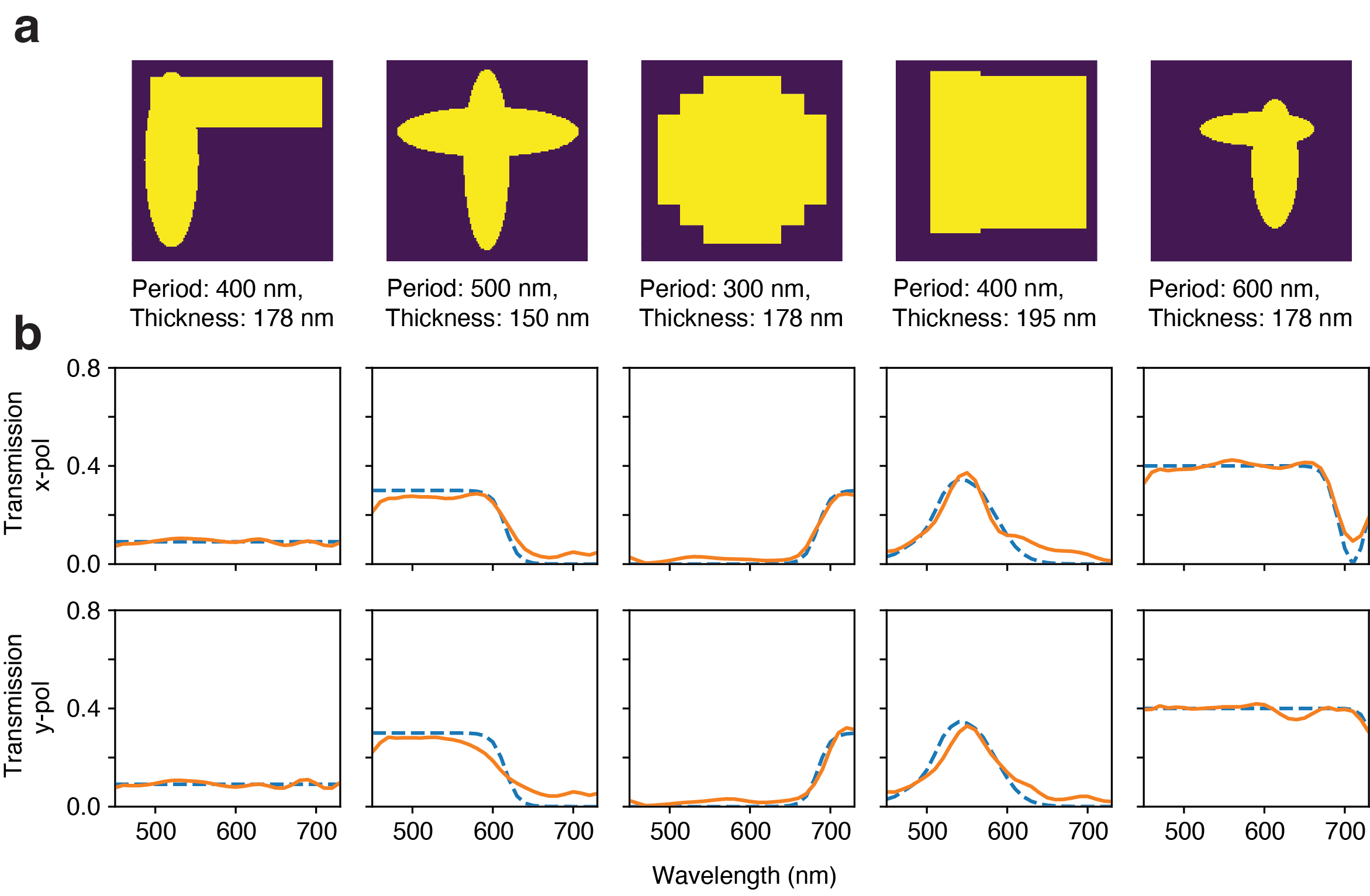}
	    \caption{\textbf{Latent space optimization results.} \textbf{a} Examples of geometries predicted by the latent space optimization. \textbf{b} Comparison between the target spectra (blue dashed lines) and simulated spectra (orange solid lines) for the predicted geometries.}
        \label{fig:optimization_res}
\end{figure}

Beyond zero-shot prediction, MOCLIP’s shared latent space enables an alternative inverse design strategy that performs optimization directly over latent vectors. This approach replaces time-intensive electrodynamic simulations of probe geometries~\cite{li2022empowering, sunae2023revisiting} with rapid inference through the geometry encoder. It substitutes a traditional spectrum-matching loss with a similarity-based loss computed between the latent vectors of the target spectrum and candidate geometries. While zero-shot prediction prioritizes high-throughput design generation, latent space optimization emphasizes high-fidelity solutions.\\ 
Figures~\ref{fig:optimization}a-e illustrate the optimization pipeline. The procedure begins by passing the target spectrum $S_t$ through the spectra encoder, resulting in its corresponding latent vector $C_t$ (Fig.~\ref{fig:optimization}a). Simultaneously, the geometry encoder processes a probe geometry $G_n$ to produce its latent vector $P_n$ (Fig.~\ref{fig:optimization}b). The same geometry generator used during the dataset creation produces these probe geometry candidates (Fig.~\ref{fig:optimization}c). The optimization routine maximizes the similarity score $C_t \cdot P_n$ by applying particle swarm optimization (PSO)~\cite{Makarenko2021} (Fig.~\ref{fig:optimization}d) to search the generator’s parameter space for the optimal geometry. This optimization continues until the algorithm finds a probe geometry whose latent representation closely matches that of the target spectrum, resulting in an optimized geometry $G_f$  (Fig.~\ref{fig:optimization}e).\\
Figure~\ref{fig:optimization}f presents the performance of latent-space optimization inverse design, evaluated on \num{500} custom target spectra previously used for library search analysis (see Supplementary Note 7). The plot shows the statistical distribution of the MSE between the target and FDTD-simulated spectra of the predicted geometries (see Methods) as a function of the number of PSO iterations. Shaded regions denote quantile bands, while dashed lines indicate the corresponding quantiles from the library search as stationary reference values.\\
The process begins with 200 randomly generated probe geometries. While library search yields better results during the initial iterations ($<5$), the optimization approach surpasses the library's performance at the 25th percentile (Q25) by the fifth iteration, improving on the most accurate library-retrieved candidates. By the 15th iteration, the optimization exceeds library performance in both mean MSE and the 75th percentile (Q75), suggesting superior refinement of typical ("median-accuracy") results. Beyond 50 iterations, the approach outperforms the library search even at the 95th percentile (Q95), demonstrating its efficacy for the most challenging target candidates. After 200 iterations, the Q95 MSE reaches $2.5 \cdot 10^{-3}$, achieving results that are competitive with, and in some cases superior to, current state-of-the-art models~\cite{liu2018generative, wang2023_inn_metasurface}. Figure~\ref{fig:optimization_res} provides representative examples of the inverse-designed geometries obtained through latent space optimization and comparisons between their FDTD-simulated spectra and the corresponding target spectra. Supplementary Note 8 presents additional experimental validation. Supplementary Table~1 compares MOCLIP against recent metasurface inverse design frameworks. Compared to published methods, MOCLIP achieves the lowest average MSE of $1.1\!\cdot\!10^{-3}$ with the fastest inference time of 500~ms while operating over the largest DoF space (42 DoFs). Optimization proceeds without forward solver calls, achieving convergence within 50 PSO iterations.

\subsection*{Information storage}
\begin{figure}[htb] \centering
	    \includegraphics[width=\linewidth]
        {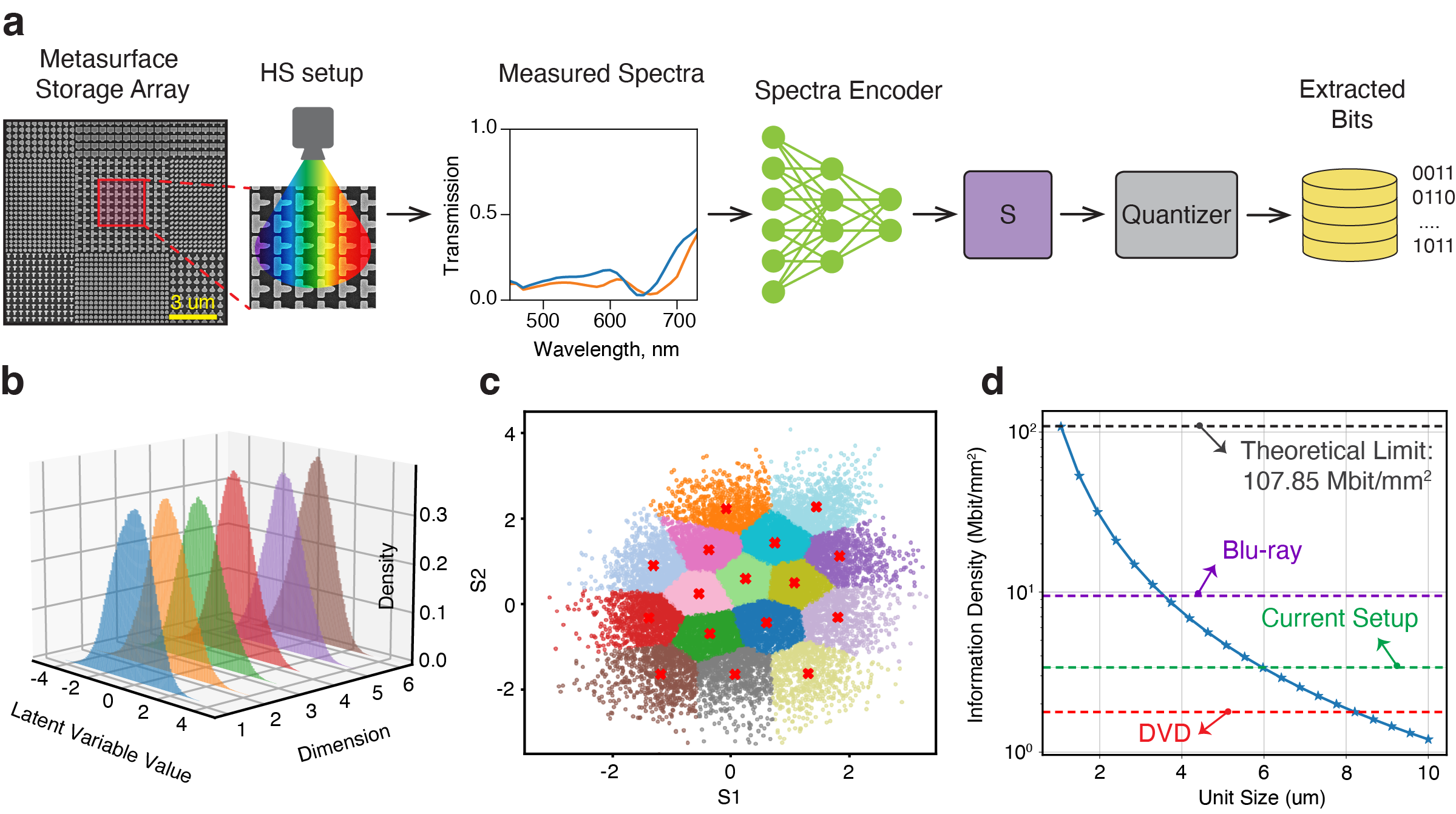}
	    \caption{\textbf{Optical storage based on latent space encoding.} \textbf{a} General schematic of optical storage utilizing a metasurface structure's latent space information. \textbf{b} Visualization of the distribution of latent vectors in their first 6 dimensions. \textbf{c} Vector quantization of the latent variables at the first 2 dimensions, where dots represent data points and red crosses are for codebook centroids. \textbf{d} Performance analysis on information density compared to DVD and Blu-ray disks. As the unit size of metasurface decreases, the theoretical density of MOCLIP optical storage outperforms Blu-ray.}
        \label{fig:storage}
\end{figure}

Metasurfaces provide a durable medium for long-term optical information storage. By exploiting the structural degrees of freedom to manipulate optical responses across the continuous spectrum, metasurfaces enable data encoding at densities beyond binary representations~\cite{zijlstra2009five}. A central challenge in such systems is retrieving the encoded information from the measured optical response, which requires classifying overlapping spectral signatures that vary due to fabrication imperfections and measurement noise. In this context, Ref.~\cite{wiecha2019storage} demonstrated that neural networks can reliably decode stored digital information from measured optical scattering spectra, achieving information densities of up to 9 bits per metasurface. However, unlike the readout process, the writing strategy still relies on multibit grid-based encoding. Here, MOCLIP extends this paradigm to a unified read–write framework by using a shared geometry–spectra latent space, enabling arbitrary information sequences to be encoded into free-form metasurface geometries and reliably recovered from their optical spectra, exploiting the design flexibility of free-form metasurfaces.\\
Figure~\ref{fig:storage}a presents an overview of MOCLIP's information storage concept. The central idea is to encode information into latent vectors that can subsequently be realized as metasurface geometries (writing process). To ensure a deterministic writing process, we leveraged the predefined candidate library provided by our dataset. We characterized the fabricated arrays using the hyperspectral transmission microscope (reading process), after which the measured transmission spectra were fed into the MOCLIP spectra encoder to obtain the latent vectors. These latent vectors are then quantized to recover the information stored in the metasurfaces.\\
Two principal error sources govern the capacity and reliability of information storage in this work: fabrication and measurement errors. Error analysis performed in Supplementary Note~3 jointly captures measurement and fabrication errors into three contributions: spatial averaging over neighboring camera pixels, spectral variations between duplicated structures, and camera noise. This combined error metric serves as the distortion value $D_i$ in the subsequent rate-distortion analysis.\\
A key requirement for latent-space information storage is that the learned representation admits an efficient quantization strategy. Figure~\ref{fig:storage}b presents the distribution of variable values in the first six latent vector dimensions. We leverage Kolmogorov–Smirnov test (K-S test)~\cite{simardComputing2011} to compare their distribution to a continuous Gaussian probability density function to better characterize and quantify the latent vectors. The resulting K-S score of \num{0.014} indicates that the 64-dimensional latent vectors closely approximate a Gaussian distribution. This finding guides the choice of a high-dimensional vector quantization (VQ) strategy, commonly adopted in modern LLM training~\cite{liuVPTQ2024}. The VQ maps the latent vectors to a discrete set of candidate vectors, namely codebook centroids, and iteratively refines them to minimize the distortion error during the mapping. Figure~\ref{fig:storage}c exemplifies the VQ strategy applied to the first two dimensions ${S_1, S_2}$ of the latent space, where $K$-means clustering constructs the mapping and locates the codebook centroids. This example applies a bit depth of \num{2} to each dimension, resulting in a codebook size of $K=16$. \\
Rate-distortion theory~\cite{coverElements2006} allows an exploration of the theoretical limit of MOCLIP's storage capacity. Equation~\ref{eq:rate_distorion} calculates the latent space's rate-distortion bound $R$, representing the maximum quantization bit depth applicable.
\begin{equation}
\label{eq:rate_distorion}
    R =\left\lfloor\sum_i\frac{1}{2} \log_2\qty(\frac{\sigma_i^2}{D_i})\right\rfloor, \quad 0 < D_i < \sigma_i^2
\end{equation}
Here, $\sigma^2_i$ represents the variance of the latent dimension $i$. Calculating $D_i$ and $\sigma_i$ on the metasurface spectra dataset results in a rate-distortion bound of $R = 126$. This value indicates that it is possible to encode a maximum of \num{126} bits of information into the 64-dimensional latent space of each metasurface. Figure~\ref{fig:storage}d plots the information density of the proposed technology against the lateral unit size of square metasurfaces. In the current experimental realization, each metasurface has a \SI{6}{\micro\meter} unit size, corresponding to an information density of \num{3.34} $\text{Mbit/mm}^2$, nearly doubling that of DVD (Fig.~\ref{fig:storage}d, red dashed line). However, reducing the metasurface sizes can yield orders of magnitude improvement of the information density. Fig.~\ref{fig:storage}d shows that MOCLIP's information storage density surpasses that of Blu-ray (purple dashed line) when the metasurface has a unit size below \SI{3.5}{\micro\meter}. The theoretical upper bound of information density with the current setup results from repeating the preceding analysis for metasurface unit sizes down to the Rayleigh criterion diffraction limit, corresponding to \SI{1.056}{\micro\meter}, and yields \num{107.85} $\text{Mbit/mm}^2$.


\section*{Discussion}
This work develops and implements MOCLIP, a framework for broadband amplitude inverse design that builds on the CLIP foundation model to learn a shared latent space for metasurface geometries and their corresponding spectra. The FM training pipeline leverages an experimentally generated dataset comprising \num{466 537} silicon-on-glass metasurface geometry-spectra pairs spanning \num{42} design degrees of freedom. The MOCLIP dataset exceeds the size of the largest previously reported metasurface–spectra dataset by more than a factor of nine and increases the supported design complexity by nearly a factor of three. Moreover, as an experimentally acquired dataset, it naturally includes information that is typically missing from simulation-based datasets, such as fabrication tolerances and measurement uncertainties. The MOCLIP experimental workflow generates an entire dataset at a rate of around \num{35} structures per minute, making this approach over \num{30} times faster than strategies based on computational electrodynamic approaches performed on supercomputers~\cite{li2022inverse}, and about 100 times higher in our own benchmark comparison (see Methods). Adopting state-of-the-art manufacturing techniques, such as immersion lithography~\cite{liLargearea2020}, can further increase this rate, enabling orders of magnitude improvements in generating robust and physically accurate datasets for nanophotonic FMs.\\
We demonstrate multiple applications of MOCLIP for inverse design via latent space optimization and zero-shot prediction. These capabilities surpass traditional reconstruction-based deep learning approaches in nanophotonics, offering up to six orders of magnitude faster inference and a tenfold improvement in accuracy~\cite{liu2018generative, zhu2024rapid},  while retaining high flexibility to incorporate fabrication-related design constraints. MOCLIP’s zero-shot prediction achieves inverse design at a speed of approximately $2 \cdot 10^5$ target spectra per second on a GPU while maintaining an MSE below $10^{-2}$, making wafer-scale, high-throughput inverse design feasible. In addition, in a CPU-based benchmark against direct library search, zero-shot prediction achieves an approximately tenfold higher throughput while preserving nearly the same accuracy.
At the same time, when higher accuracy is required, MOCLIP further supports optimization-based inverse design in the learned latent space. In this regime, the method refines the initial zero-shot solutions and surpasses direct library search, reducing the Q95 MSE to $2.5 \cdot 10^{-3}$ after 200 PSO iterations. This demonstrates that MOCLIP combines rapid large-scale screening with accurate continuous refinement within a single framework.\\
The present framework generalizes to phase information by conventional holography-based techniques~\cite{ref1, ref2, ref3}. The hyperspectral transmission microscope can be extended to a holographic configuration by introducing a reference arm of matched path length, enabling simultaneous retrieval of amplitude and phase across the full spectral range. On the software side, the network architecture and training pipeline remain unchanged, leveraging the proposed CLIP structure. Such modularity is a core advantage of CLIP-based models, which naturally support the incorporation of multiple input modalities within a shared latent space~\cite{moro2023multimodal}.\\
We concluded this work by demonstrating an application of MOCLIP for optical information storage using a library-assisted latent vector encoding scheme. Our approach leverages the shared geometry-spectra latent space to encode information using a library of predefined candidate structures, with VQ employed to model the distribution of latent vectors and optimize the quantization strategy. With the current experimental realization, MOCLIP achieves an information density of \num{3.34} $\text{Mbit/mm}^2$, comparing favorably to last-generation optical storage devices. This concept can be extended to outperform the state-of-the-art by shrinking the metasurface unit size, with a theoretical upper bound of \num{0.1} $\text{Gbit/mm}^2$ at a metasurface unit size of \SI{1.056}{\micro\meter}, set by the diffraction limit of the current optical system. Future work can generalize this idea to free-form design with continuous inverse mapping from arbitrary bit sequences, opening the path toward complete read/write optical storage operations.

\section*{Methods}

\subsection*{Geometries generation}
The generation algorithm uses a Python-based software tool that stochastically populates a square cell with a fixed lattice period with a user-specified number of primitives (rectangles, ellipses, and rings). For each primitive, the algorithm samples the center position and in-plane dimensions (for rings, both outer and inner dimensions), accepting the draw only if the feature lies entirely within the cell. To satisfy fabrication constraints, the algorithm enforces limits on each shape's dimensions, the spacing between features, and the minimum feature size. It requires every shape dimension to be at least \SI{100}{\nano\meter}, and the feature size and inter-feature gap to be no smaller than \SI{50}{\nano\meter}. The software samples each design on a discrete grid with lattice periods of \SIlist{300;400;500;600}{\nano\meter} and silicon thicknesses of \SIlist{150;178;195}{\nano\meter}. To keep the design space within practical limits, it allows at most four primitives per unit cell, resulting in 42 independent geometric degrees of freedom. Within these bounds, the software generated \num{546 120} unique metasurface designs.

\subsection*{Fabrication} 
The process begins with cleaning a fused silica wafer using acetone and isopropyl alcohol, followed by the deposition of a uniform amorphous silicon layer via low-pressure chemical vapor deposition (LPCVD). The thickness of the silicon layer is precisely controlled using ellipsometry (UVISEL Plus, HORIBA). The wafer is then diced into \SI{15}{mm} wide and \SI{500}{\micro\meter} thick pieces to serve as base samples. A positive electron beam resist, AR-P 6200.04 (Allresist), is spin-coated on the samples at \SI{6000}{RPM} for \SI{60}{s}, followed by a bake on a hotplate for \SI{90}{s} at \SI{150}{\celsius}. Subsequently, a conductive polymer, AR-PC 5090.02 (Allresist), is spin-coated at \SI{4000}{RPM} for \SI{60}{s} and baked at \SI{100}{\celsius} for \SI{1}{\minute}. Metasurface patterns are written using a JEOL JBX-6300FS electron beam lithography system at \SI{100}{\kilo\volt}. After patterning, the conductive polymer is removed by submerging the sample in deionized water for \SI{60}{s}. The resist is developed using AR 600-546 (Allresist) for \SI{60}{s}, followed by O-Xylene for \SI{7}{s} to sharpen feature edges. A \SI{22}{nm} chromium layer is deposited via electron beam evaporation to form an etch mask. Unwanted chromium is lifted off by submerging the sample in AR 600-71 (Allresist) for 10 minutes, followed by one minute of sonication. Reactive ion etching with a $\text{CHF}_3$ and $\text{SF}_6$ mixture is used to remove unprotected silicon and expose the underlying substrate. Finally, the chromium mask is removed by submerging the sample in a mixture of perchloric acid and ceric ammonium nitrate (TechniEtch Cr01, MicroChemicals) for \SI{60}{s}.

\subsection*{Foundation model training}
The dataset is divided into three subsets: training (\num{373 230} samples; 80\%), validation (\num{46 654} samples; 10\%), and test (\num{46 653} samples; 10\%). The model is trained exclusively on the training set, using the validation to monitor performance at the end of each epoch and guide model selection. The test set was reserved to evaluate the final model performance after the training.
Within the full dataset, geometries contain 2 shapes in 70\% of samples, 3 shapes in 20\%, and 1 or 4 shapes in 5\% each. Periods and shape types are distributed uniformly, and thickness is approximately uniform ($33\% \pm 3\%$ per bin). All samples were verified to be unique using hashing.
\\
The model is trained for 140 epochs using a three-stage learning rate schedule: $1 \cdot 10^{-4}$ for epochs 1--60, $5 \cdot 10^{-5}$ for epochs 61--100, and $3 \cdot 10^{-5}$ for epochs 101--140. Adam optimizer~\cite{kingma2014adam} updates the model parameters. The network contains approximately 11.8 million trainable parameters. The Supplementary Materials further detail the model architecture. We trained the model with three different seed values and monitored Top-k accuracy on the validation set. The results showed less than 0.2\% accuracy difference for the Top-1, Top-5, and Top-10 metrics, so we arbitrarily selected the first model. All training and evaluation are conducted on a workstation equipped with an AMD Ryzen Threadripper PRO 5955WX 16-core processor, 128 GB of RAM, and an NVIDIA GeForce RTX 4090 GPU.\\

\subsection*{Finite-Difference Time-Domain validation and speed benchmark}
To validate the inverse-design results from latent-space optimization, we performed full-wave FDTD simulations using our custom solver for the final geometries generated by the particle swarm optimizer. To minimize the impact of numerical error on this validation, we first assessed spectral convergence with respect to grid resolution. Specifically, we computed the MSE between spectra obtained at different grid resolutions and a reference spectrum calculated with a \num{2.5} nm grid resolution (Fig.~S3a). We also measured the computation time for each grid resolution (Fig.~S3b). These benchmarks were performed on an HPE Cray EX system using 96 CPU cores. We defined one full simulation as the combination of a free-space reference simulation and a structure simulation for a single polarization state. The results indicate that a \num{5} nm grid resolution provides a suitable compromise between accuracy and computational cost: the corresponding error, $\mathrm{MSE}=10^{-4}$, is one order of magnitude smaller than the experimental measurement error and therefore has a negligible effect on the validation results. At this resolution, the computation time in our configuration was \num{3} min. Figure~S3c compares the FDTD and experimental results, showing that the discrepancy remains within the measurement uncertainty.

\section*{Data availability}
A subset of the dataset comprising 10,000 samples is openly available in the GitHub repository at \url{https://github.com/RodionovSA/MOCLIP_release} and archived on Figshare~\cite{moclip_data}. The full dataset is available from the corresponding author upon request.

\section*{Code availability}
The full source code used in this study is openly available in a GitHub repository at \url{https://github.com/RodionovSA/MOCLIP_release} and archived on Zenodo~\cite{moclip_code}.

\printbibliography
 
\section*{Acknowledgements}
We performed the metasurface fabrication and scanning electron microscopy analysis at the KAUST Nanofabrication Core Lab and Imaging and Characterization Core Lab, respectively. We are grateful to the Core Labs staff for their consistent technical support, and we extend our special thanks to Dr. Camelia Florica for her essential assistance in the fabrication process optimization.

\section*{Author Contributions}
A.F. supervised the work. A.B.L., M.M., and F.G. conceptualized the idea and acquired preliminary results. S.R. developed the metasurface geometry generator algorithm and generated the dataset of metasurface geometries. S.R. and A.B.L. fabricated the samples, built the automated hyperspectral microscopy setup, and performed hyperspectral measurements. S.R. processed the hyperspectral measurement data and built the final dataset. S.R. and M.M. developed the MOCLIP model. S.R. trained and tested the model. S.R. developed the algorithms for zero-shot prediction, library-based search, and latent space-based optimization. Q.W. developed the information storage idea and performed memory capacity evaluation. S.R., A.B.L., Q.W., and A.F. wrote the manuscript.

\section*{Competing Interests}
The authors declare no competing interests.


\clearpage

\end{document}


\title{Supplementary Materials -- MOCLIP: A Foundation Model for Large-Scale Nanophotonic Inverse Design}

\maketitle

\begin{affiliations}
\item{PRIMALIGHT, Faculty of Electrical Engineering; Applied Mathematics and
    Computational Science, King Abdullah University of Science and Technology,
    Thuwal 23955-6900, Saudi Arabia
    }
\item{
    Sber AI Lab, Moscow, Russia\\
    $^\dag$ These authors contributed equally.\\
    $^*$ Corresponding author. Email: andrea.fratalocchi@kaust.edu.sa.
    }
\end{affiliations}


\section*{Supplementary Note 1. Fabrication details}
To account for the dependence of metasurface transmission spectra on thickness, we designed metasurface geometries with three thickness values: \num{150}~nm, \num{175}~nm, and \num{200}~nm (see Methods of the main text). To fabricate samples at each target thickness, we divided them into three groups, each receiving an independently deposited amorphous silicon film. We validated the thickness values by ellipsometry, yielding final measured values of \num{150}~nm, \num{178}~nm, and \num{195}~nm. MOCLIP's capabilities are also applicable to the design of multiple thickness geometries on a single substrate, which can be achieved through techniques such as grayscale electron-beam lithography~\cite{shanks2025metasurfaces} or multiphoton lithography~\cite{vanmol2024fabrication}, allowing improved performance for broadband devices. \\
We distributed all \num{546120} designs among three thickness groups as follows: \num{136530} designs for \num{150}~nm, \num{204795} for \num{178}~nm, and \num{204795} for \num{195}~nm. Given the chosen metasurface unit cell size of $6~\mu\text{m} \times 6~\mu\text{m}$ and a camera sensor size of $14.13~\text{mm} \times 10.35~\text{mm}$ at $10\times$ magnification, \num{22755} metasurfaces fit within a single field of view with adequate spacing. We therefore partitioned the dataset into arrays of \num{22755} metasurfaces, with 6 such arrays printed per fused silica substrate -- \num{3} arrays of unique designs and \num{3} duplicate arrays for cross-validation. This yielded \num{2} substrates at \num{150}~nm, \num{3} at \num{178}~nm, and \num{3} at \num{195}~nm. In total, we fabricated $\num{22755} \cdot 6 \cdot 8 = \num{1092240}$ metasurfaces, of which half comprised unique designs. Following cross-validation, the final dataset consists of \num{466537} samples. 

\section*{Supplementary Note 2. Experimental setup details}
In this work, we performed hyperspectral imaging using a spectral-scanning technique~\cite{Gao2015}. In contrast to spatial scanning techniques, such as line~\cite{Cutler2013} or point scanning~\cite{Sinclair2006}, our approach captures 2D monochrome images for each wavelength independently. This method allows us to exploit the full resolution of CMOS cameras (12.3MP with \num{3.35} $\mu$m pixel size in our case) when recording metasurface responses at each spectral band. We achieve spectral control of the illumination by using a tunable bandpass filter, as shown in Fig.~3e of the main text.\\
Specifically, broadband light from a xenon lamp (HPX-2000, Ocean Optics) is collimated by a $10\times$ objective (RMS10X, Thorlabs) and linearly polarized using a Glan–Taylor prism (DGL10, Thorlabs)  before passing through a liquid crystal tunable bandpass filter (K2VB1, Thorlabs). We drive the filter at its minimum bandwidth (\SI{10}{nm} FWHM at \SI{550}{nm}) and step it from \SIrange{450}{ 730}{nm} in \SI{10}{nm} increments. We use a half-wave plate (AHWP10M-600, Thorlabs) to set the required x- or y-polarisation. We then expand the beam two-fold with a 4-f telescope (lenses LA1131-AB and LA1433-AB, Thorlabs) and spatially clean it with an adjustable aperture. We divide the beam with a non-polarising beamsplitter (BSW26R, Thorlabs). One arm images onto a reference CMOS camera (LP126MU, Thorlabs) to monitor the source intensity, and the other continues to the sample arm of the microscope. We trim the beam diameter with a field-stop aperture and then focus it using a plano-convex lens (LA1509-AB, Thorlabs) onto the back focal plane of a $10\times$ long-working-distance objective (MY10X-823, Thorlabs), establishing Köhler illumination on the sample. A second $10\times$ long-working-distance objective collects the transmitted light and relays it through a \SI{200}{mm} tube lens (TTL200-S8, Thorlabs) onto the imaging CMOS camera (CS126MU, Thorlabs), yielding high-contrast metasurface images for every wavelength–polarization setting.\\
We mount the sample on a motorized pitch–yaw platform (PY004Z9, Thorlabs) stacked on a three-axis linear stage (PT3Z9, Thorlabs), providing five motorized degrees of freedom ($X$, $Y$, $Z$, $\theta_x$, $\theta_y$) for precise positioning. A precision rotation mount (CRM1PT, Thorlabs) adds manual rotation about the optical axis, allowing the camera sensor to register the metasurface array angularly. The stage’s travel range is large enough to completely translate the sample from the objective’s field of view, enabling fully automated measurement cycles. 

\section*{Supplementary Note 3. Measurement error analysis}
\begin{figure}[h] \centering
	    \includegraphics[width=\linewidth]{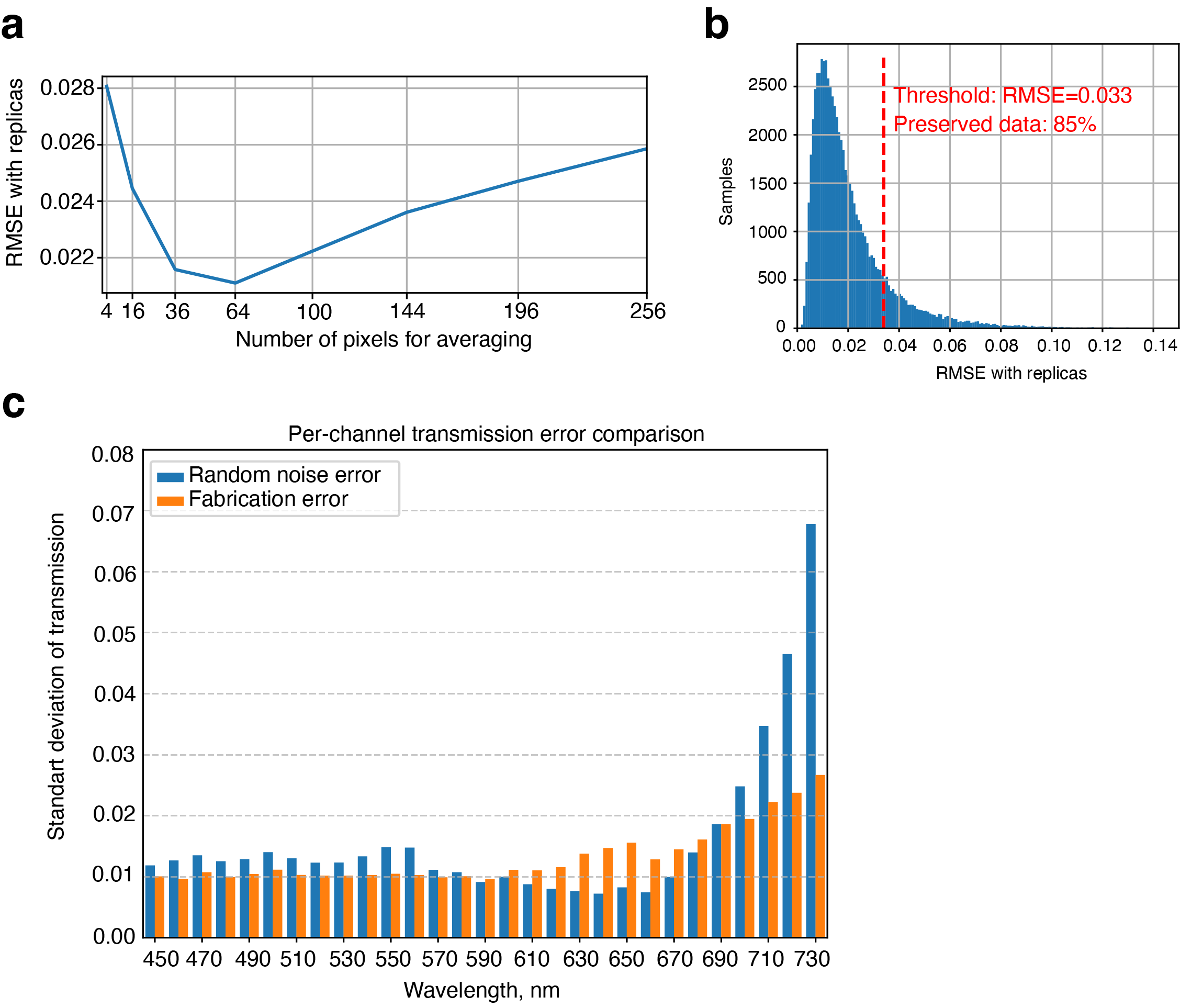}
	    \caption{\textbf{Measurement error analysis.} \textbf{a.} RMSE between original and replicated metasurface spectra as a function of the number of averaged pixels. Averaging over the central \(8 \times 8\) (\num{64}) pixels yields the lowest deviation of \num{0.021}. \textbf{b.} RMSE distribution between original and replicated spectra for quality control. \textbf{c.} Wavelength-dependent random and fabrication errors, \(\sigma_{\mathrm{rand}}\) and \(\sigma_{\mathrm{fab}}\).  
        \label{fig:error}}
\end{figure}

First, we analyze the optimal number of pixels to average for each metasurface, balancing the trade-off between camera noise and boundary effects. From the noise reduction perspective, averaging over a larger number of pixels minimizes random fluctuations and improves signal stability. However, including pixels near the edges of each metasurface can introduce errors due to optical cross-talk or fabrication imperfections from neighboring structures. To identify the optimal averaging area, we conduct an error analysis by comparing spectra from original structures and their exact replicas (fabricated and measured independently) while varying the number of averaged pixels. We find that averaging over the central \num{64} pixels out of \num{289} provides the best trade-off, yielding the lowest mean spectral deviation of \num{0.021} (Fig.~\ref{fig:error}a).

Next, we perform cross-validation on the spectral dataset by comparing each spectrum with that of its corresponding replica, following the same procedure used in the pixel-averaging analysis. Figure~\ref{fig:error}b shows the results, quantified using root mean squared error (RMSE). To mitigate the influence of fabrication defects, surface contamination, and measurement inconsistencies, we apply a quality control threshold that excludes samples with a mean squared error (MSE) greater than $1 \cdot 10^{-3}$ from the dataset. This filtering step retains approximately $85\%$ of the dataset. 

Finally, we quantify the measurement uncertainty associated with the validated spectra. Each spectrum carries three independent 1-$\sigma$ uncertainties:  
(i) camera shot and read noise after five-frame and \(8\times 8\) patch averaging,  
(ii) pixel-to-pixel fabrication variability inside the \(8\times 8\) analysis patch,  
(iii) the wavelength setting accuracy of the liquid-crystal tunable filter (LCTF).

For every structure \(j\) and wavelength slice \(i\) we first propagate the
per-pixel noise~$\sigma_{T,m}$. The random error of the pixel average is
\begin{equation}
  \sigma_{\mathrm{rand}}(i,j)=
  \frac{\sqrt{\displaystyle\sum_{m=1}^{N_p}\sigma_{T,m}^2}}{N_p},
  \label{eq:sigma_rand}
\end{equation}

where \(N_p=64\).
The fabrication spread is
\begin{equation}
  \sigma_{\mathrm{fab}}(i,j)=
  \sqrt{\frac{1}{N_p-1}\sum_{m=1}^{N_p}\!\bigl[T_m(i,j)-\bar T(i,j)\bigr]^2},
  \label{eq:sigma_fab}
\end{equation}

with \(\bar T\) being the mean of the \(8\times 8\) patch. Fig.~\ref{fig:error}c shows the wavelength-dependent variation of both error components, with $\sigma_{\mathrm{rand}}$ and $\sigma_{\mathrm{fab}}$ averaged over all structures and plotted as functions of wavelength. The random and fabrication uncertainties combine in quadrature as
\begin{equation}
\sigma_{\mathrm{tot}}(i,j) =
\sqrt{\sigma_{\mathrm{rand}}^2 + \sigma_{\mathrm{fab}}^2},
\label{eq:sigma_tot}
\end{equation}
yielding structure and wavelength-averaged values of $\sigma_{\mathrm{rand}}$ = \num{0.016}, $\sigma_{\mathrm{fab}}$ = \num{0.013}, and $\sigma_{\mathrm{tot}}$ = \num{0.026}. In addition, the liquid-crystal tunable filter (LCTF) introduces a separate systematic wavelength uncertainty of $\pm 0.5\;\mathrm{nm}$, specified by the manufacturer. 

\section*{Supplementary Note 4. MOCLIP architecture}
\begin{figure}[htb] \centering
	    \includegraphics[width=\linewidth]{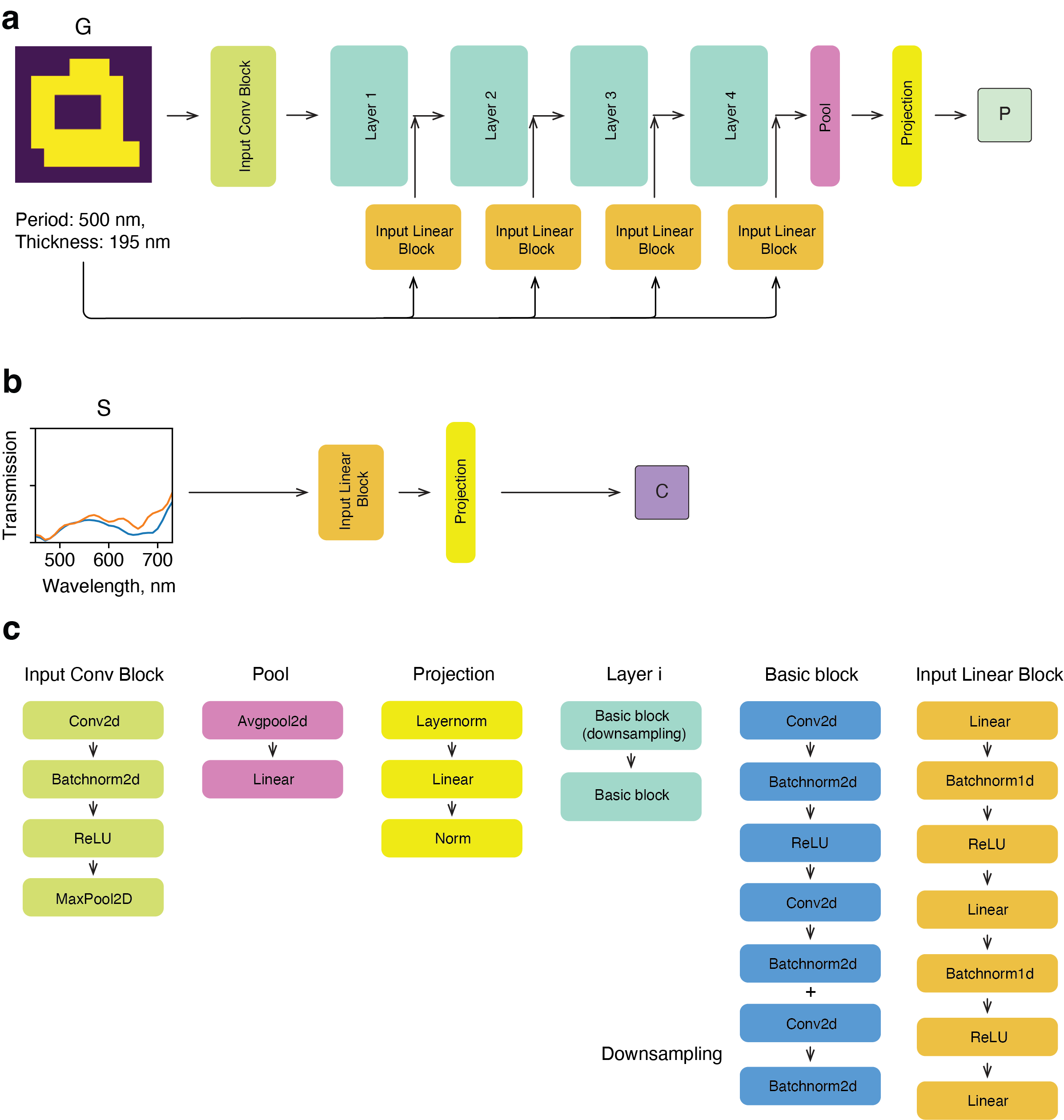}
	    \caption{\textbf{MOCLIP architecture.} 
\textbf{a.} Geometry encoder architecture based on ResNet. \textbf{b.} Spectra encoder architecture based on MLP. \textbf{c.} Structure of individual network blocks.}
     \label{fig:moclip_struct}
\end{figure}

Figure~\ref{fig:moclip_struct} illustrates the architecture of MOCLIP’s geometry and spectra encoders. In the geometry encoder (Fig.~\ref{fig:moclip_struct}a), the input geometry $\mathbf{G}$ is represented as a $128 \times 128$ binary image, accompanied by its period and thickness values. The image is processed through the Input Convolution Block following the standard ResNet pipeline, while the period and thickness parameters are transformed via Input Linear Blocks, tiled to match the spatial dimensions of the feature maps, and added after each ResNet Layer Block. This design ensures that these scalar parameters remain integrated throughout processing, guiding the image-based feature extraction. Finally, a Projection Layer maps the combined features to a \num{64}-dimensional latent vector $\mathbf{P}$.  

The spectra encoder (Fig.~\ref{fig:moclip_struct}b) takes as input a \num{58}-dimensional concatenated transmission spectrum for $x$- and $y$-polarizations, $\mathbf{S}$, and processes it through an MLP-based architecture. It begins with an Input Linear Block, followed by a Projection Layer that outputs a \num{64}-dimensional latent vector $\mathbf{C}$, matching the geometry encoder’s latent space. The structures of all encoder blocks are detailed in Fig.~\ref{fig:moclip_struct}c.


\section*{Supplementary Note 5. Zero shot prediction of spectral response}
Figures~\ref{fig:zero-shot}a-c illustrate MOCLIP's zero-shot prediction process of spectral response of a target geometry. The process starts by encoding the target geometry $\mathbf{G}_t$ with the geometry encoder to produce its latent-space representation $\mathbf{P}_t$ (Fig.~\ref{fig:zero-shot}a). Next, MOCLIP generates a set of probe spectra $\mathbf{S}_n$ (Fig.~\ref{fig:zero-shot}b) and encodes each through the spectra encoder to obtain their corresponding latent-space representations $\mathbf{C}_n$. By computing the cosine similarity between the geometry's latent vector and each of the probe spectra latent vectors, MOCLIP generates a score vector indicating how closely each probe spectrum matches the target geometry (Fig.~\ref{fig:zero-shot}c). MOCLIP then selects the best-rated spectrum (Fig.~\ref{fig:zero-shot}c, yellow cell) as its prediction. Figures~\ref{fig:zero-shot}d–e present the Top-$k$ prediction accuracy and precision for zero-shot spectra prediction using the database generation approach, respectively.\\


\section*{Supplementary Note 6. One-to-Many Mapping in Inverse Design}
Unlike generative inverse-design models, MOCLIP does not learn a direct mapping from spectra to geometry. It is therefore not inherently affected by the one-to-many mapping phenomenon, in which multiple distinct geometries can produce nearly identical optical responses. Instead, MOCLIP learns a shared latent space for geometries and spectra through contrastive training, in which matched geometry--spectrum pairs are close to each other while mismatched pairs are pushed apart. We then perform inverse design directly in this latent space by identifying geometries whose latent vectors are most similar to the target spectral embedding, as measured by cosine similarity. Because the method retrieves any geometry that is highly consistent with the target response, rather than regressing toward a single geometry label, it naturally accommodates multiple valid solutions.\\
Figures~\ref{fig:nonuniq1} and~\ref{fig:nonuniq2} illustrate this behavior, in which MOCLIP retrieves the top-ranked geometry candidates by latent space similarity for a given target spectrum. While the highest-ranked candidate may coincide with the exact target geometry, lower-ranked candidates are generally distinct geometries that nevertheless produce spectra close to the target response, as confirmed by the reported geometry--geometry, spectrum--spectrum, and geometry--spectrum similarity scores, as well as spectral MSE. This result directly demonstrates that MOCLIP does not collapse multiple valid solutions into a single prediction, but instead identifies a set of geometrically distinct yet optically consistent candidates.


\section*{Supplementary Note 7. Custom target spectra generation}
To evaluate the inverse design framework across a diverse range of user-defined requirements, we curated a dataset of \num{500} artificial transmission spectra. We categorized these targets into four primary groups: neutral-density filters, broadband polarizers, and polarization-dependent broadband bandpass and inverted bandpass filters, which are defined as polarization-dependent color filters. By systematically varying parameters such as amplitude, bandwidth, and plateau levels, we ensured a broad representation of spectral profiles (see Figure 8 of the main text for representative examples). \\
We note that for high-absorption targets, such as polarizers and neutral-density filters, achieving precise results is challenging due to the inherent measurement error of MSE = $10^{-3}$ (see Supplementary Note 3). This error forms a noise floor that prevents accurate evaluation of candidates with transmission below 3\%, as the experimental fluctuations obscure the true performance of these high-extinction geometries. 

\section*{Supplementary Note 8. Experimental validation of design by latent space optimization}
We perform additional experimental validation of inverse-designed structures for three target spectral designs randomly selected from the set of \num{500} artificially generated spectra (see Methods). For each target, we obtain the corresponding structure after \num{200} PSO iterations. Figure~\ref{fig:optim} compares the desired spectral response, simulated spectra, and measured spectra for manufactured samples on Si nanostructures with a thickness of \SI{150}{\nano\meter}. Figure~\ref{fig:optim}b shows binary maps depicting the geometry of the basic nanostructure unit cell for each design. The simulated spectra exhibit MSE values of $2 \cdot 10^{-4}$, $9.3 \cdot 10^{-4}$, and $1.6 \cdot 10^{-3}$ with respect to the targets, while the measured spectra yield $5 \cdot 10^{-4}$, $1.5 \cdot 10^{-3}$, and $2.1 \cdot 10^{-3}$, respectively. The simulated MSE combines the inverse design error with the numerical error of the FDTD solver ($\approx 1 \cdot 10^{-4}$); isolating the latter indicates a mean inverse design error of $\approx 8 \cdot 10^{-4}$ across the three structures. The measured MSE additionally includes the combined measurement and fabrication error, which for an independent per-channel noise of $\sigma = 0.026$ (see Supplementary Note 3) contributes $\sigma^2 \approx 6.8 \cdot 10^{-4}$ to the MSE. Isolating this contribution from the mean measured MSE gives a consistent inverse design error of $\approx 7 \cdot 10^{-4}$. Both estimates agree with the mean MSE of $1.1 \cdot 10^{-3}$ obtained from FDTD validation across \num{500} custom target spectra, confirming the accuracy of the latent-space optimization approach. Figure~\ref{fig:optim}c shows SEM images of the fabricated structures.


\section*{Supplementary Note 9. Library search scaling}
We evaluated the scaling of library search using MOCLIP in the latent space and exact MSE search in the spectral space. Figure~\ref{fig:library_search}a shows the resulting scaling, where dots mark the \num{0.25}, \num{0.5}, and \num{1.0} fractions of the dataset and solid lines indicate their linear fits. At a dataset size of $10^9$, the search time per target is \num{3.380} seconds for MOCLIP versus \num{48.260} seconds for exact search, revealing a 14$\times$ performance gap in favor of MOCLIP. \\
MOCLIP's enhanced performance leverages the main difference between the two search strategies: the choice of the similarity metric and its computational implementation. Latent space search relies on dot products between normalized embeddings and selects the library entry with the highest similarity score. In contrast, a direct library search computes the MSE between each library spectrum and the target and selects the entry with the smallest error. Beyond a dataset size of $10^5$, the embedding-based formulation becomes increasingly advantageous, as modern CPU and GPU architectures are highly optimized for matrix-vector products.\\
We further accelerated direct library search in spectral space by employing an approximate nearest neighbor search based on the Inverted File (IVF) index~\cite{jegou2010product}. We applied acceleration only to the MSE-based search, while keeping the MOCLIP latent space unaltered. Figure~\ref{fig:library_search}a shows the results. IVF reduces the search time per target by approximately 2$\times$, reaching \num{22.770} seconds at a dataset size of $10^9$, around \num{7} times slower than MOCLIP. Figure~\ref{fig:library_search}b evaluates the loss of accuracy of the acceleration, comparing the 95th percentile of the Top-1 MSE of retrieved candidates across methods for different dataset fractions. The results show that the IVF accuracy is approximately $25\%$ lower than that of MOCLIP.
 

\section*{Supplementary Note 10. Comparative Analysis of Inverse Design Methods}
Supplementary Table~\ref{tab:bench} compares MOCLIP against recent metasurface inverse design frameworks across degrees of freedom, optimization iterations, accuracy, inference time, and computational platform.

\begin{table}[h]
\centering
\footnotesize
\setlength{\tabcolsep}{4pt}
\caption{Comparative Analysis of Inverse Design Methods}
\label{tab:bench}
\begin{tabularx}{\linewidth}{c l X c c c c c}
\toprule
\textbf{ID} & \textbf{Ref} & \textbf{Inverse Design Approach} & \textbf{DoFs} & \textbf{Iterations} & \textbf{MSE} & \textbf{Inf. Time} & \textbf{Platform} \\
\midrule
1 & \cite{panda2020robust}      & RCWA + GA                              & 16 & 70  & $3.7\!\cdot\!10^{-2}$ & 2.5 days & CPU \\
2 & \cite{desouza2023backpropagation} & Library + Surrogate + Grad.\ Opt. & 4  & 100 & $10^{-2}$              & 35 s     & CPU \\
3 & \cite{tanriover2022deep}    & Surrogate + GA                         & 19 & 150 & $2\!\cdot\!10^{-3}$   & N/R      & GPU \\
4 & \cite{deng2021neural}       & Surrogate + Grad.\ Opt.                         & 14 & 1000   & $1.2\!\cdot\!10^{-3}$ & 1 min    & GPU \\
5 & \cite{wang2023end}          & INN                                    & 6  & 1   & $6\!\cdot\!10^{-3}$   & 10 s     & GPU \\
6 & This work                   & MOCLIP + PSO                           & 42 & 50  & $1.1\!\cdot\!10^{-3}$ & 500 ms   & GPU \\
\bottomrule
\end{tabularx}
\vspace{4pt}

\noindent
\begin{minipage}{\linewidth}
\raggedright
\scriptsize
\textit{Abbreviations:} GA, genetic algorithm; Grad.\ Opt., gradient-based optimization; INN, invertible neural network; PSO, particle swarm optimization; N/R, not reported.
\end{minipage}
\end{table}
 
\printbibliography


\begin{figure}[htb] \centering
	    \includegraphics[width=\linewidth]{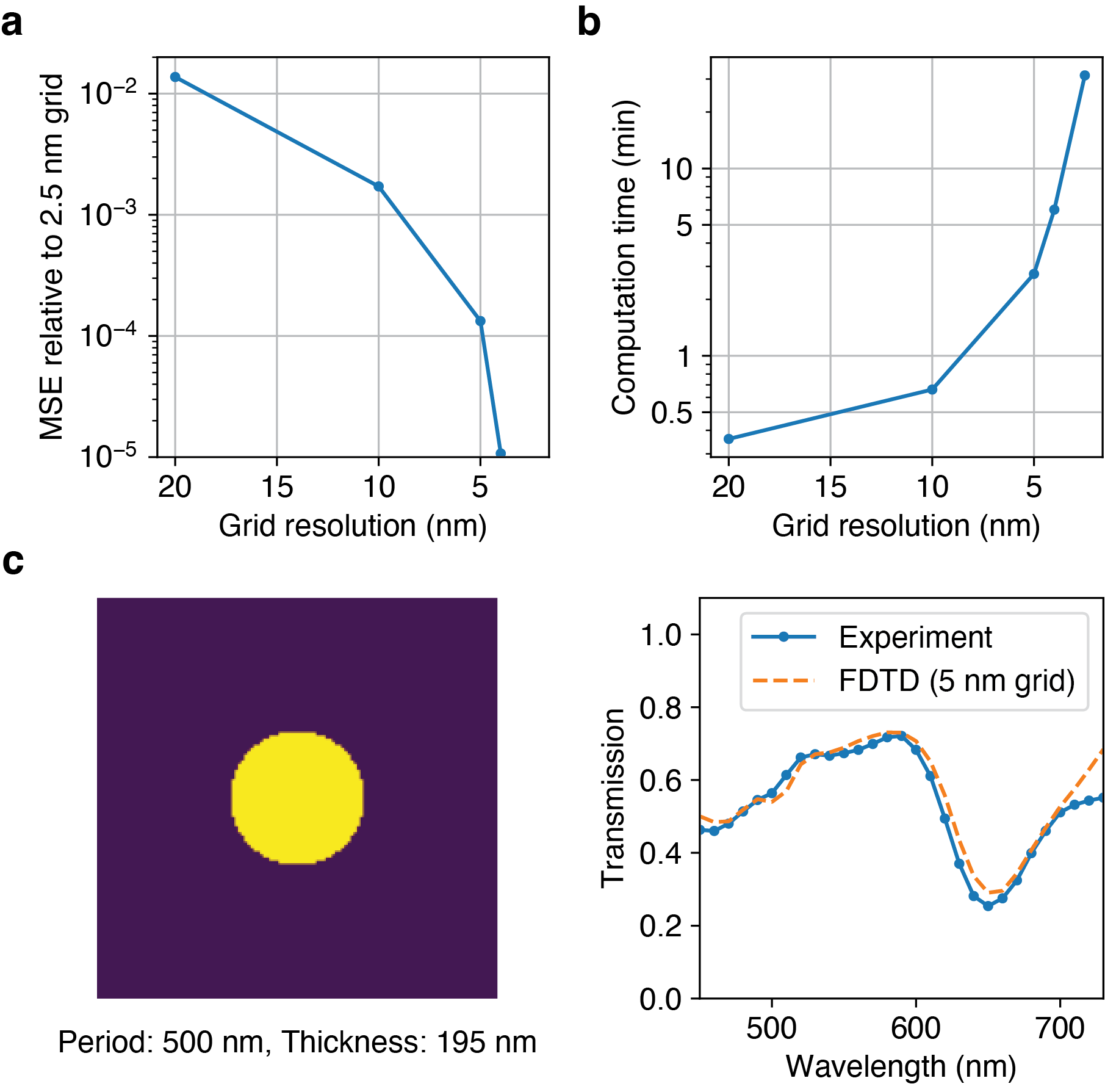}
	    \caption{\textbf{FDTD results.} 
    \textbf{a.} Convergence of the FDTD spectra with respect to grid resolution.
    \textbf{b.} FDTD computation time as a function of grid resolution.
    \textbf{c.} Comparison between FDTD and experimental results: the left panel shows the geometry, and the right panel compares the corresponding spectra.}
    \label{fig:speed}
\end{figure}

\begin{figure}[htb] \centering
	    \includegraphics[width=\linewidth]{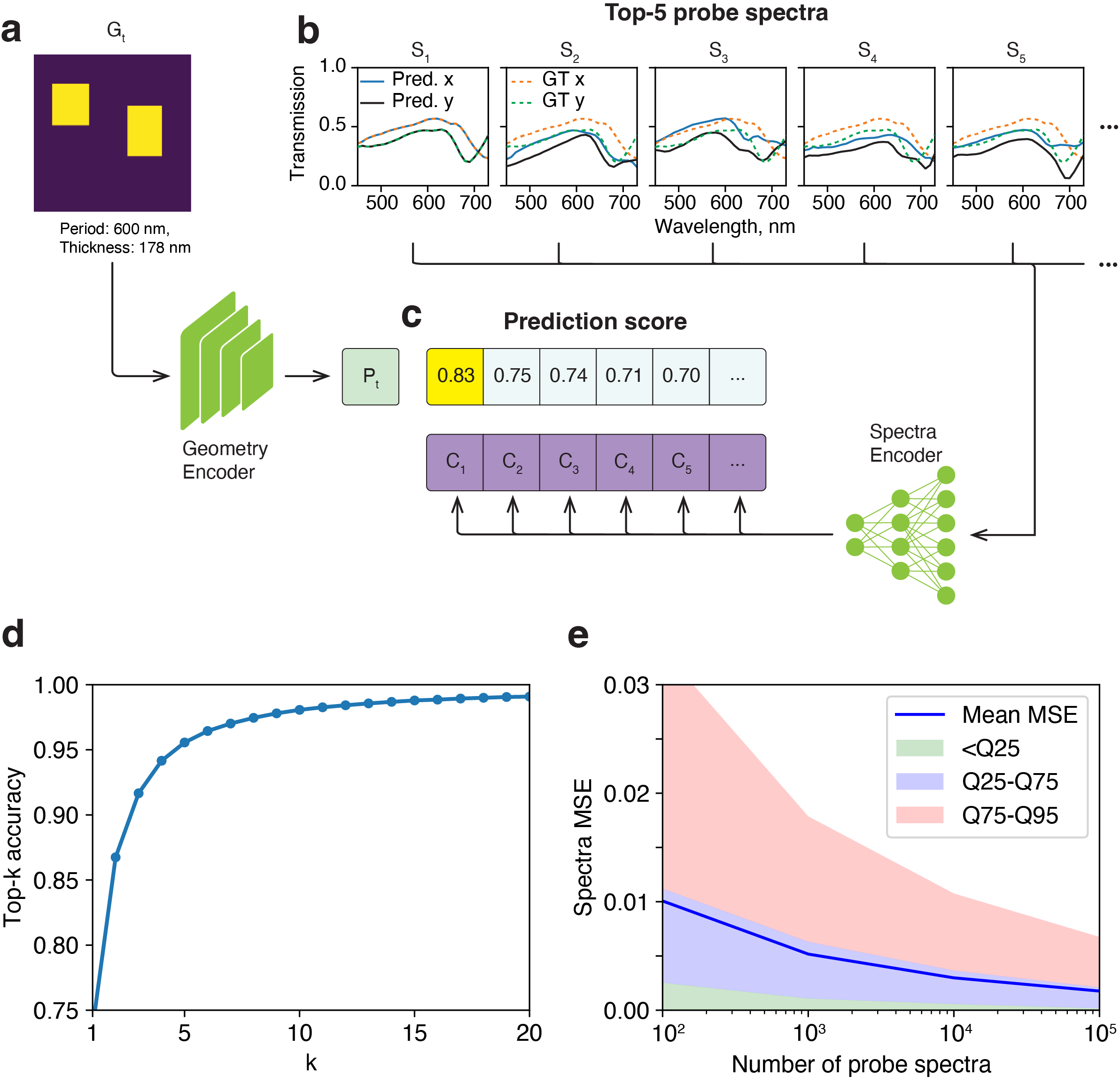}
	    \caption{\textbf{Zero-shot prediction.} 
\textbf{a.} Target geometry encoding into a latent vector. \textbf{b.} Encoding of probe spectra, including the ground truth, into latent vectors. \textbf{c.} Similarity score vector between the target geometry and the candidate spectra latent vectors. The yellow cell indicates the best matching score for the predicted spectrum. \textbf{d.} Top-k accuracy for the test dataset. \textbf{e.} Statistical distribution of the MSE between the spectra corresponding to the target geometry and predicted spectra as a function of the number of probe spectra. The shaded bands indicate quantile regions, and the solid line indicates the mean.}
     \label{fig:zero-shot}
\end{figure}

\begin{figure}[!htbp] \centering
\includegraphics[width=\linewidth,height=0.78\textheight,keepaspectratio]{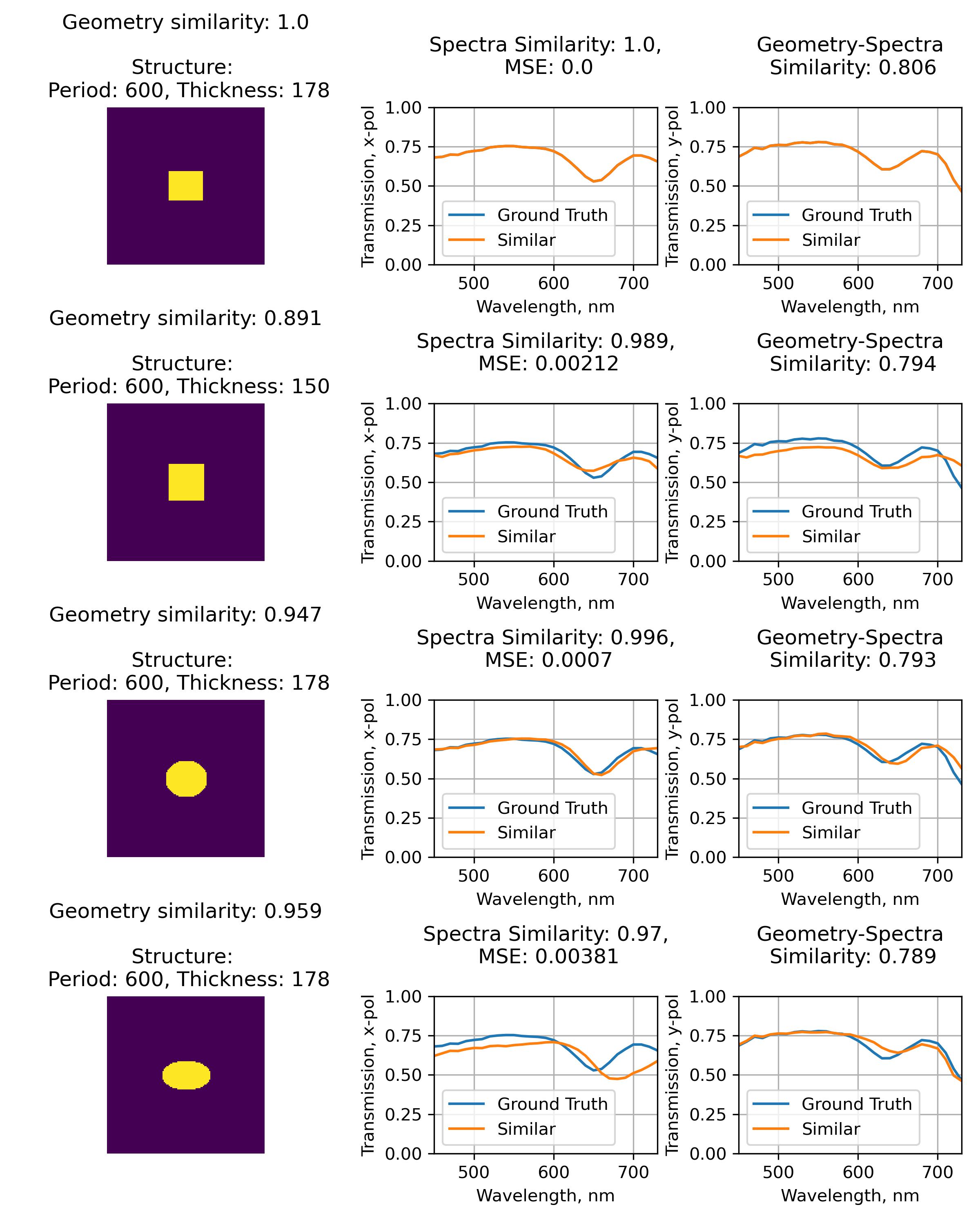}
	    \caption{\textbf{Non-uniqueness analysis for a simple geometry.} }
     \label{fig:nonuniq1}
\end{figure}

\begin{figure}[!htbp] \centering
\includegraphics[width=\linewidth,height=0.78\textheight,keepaspectratio]{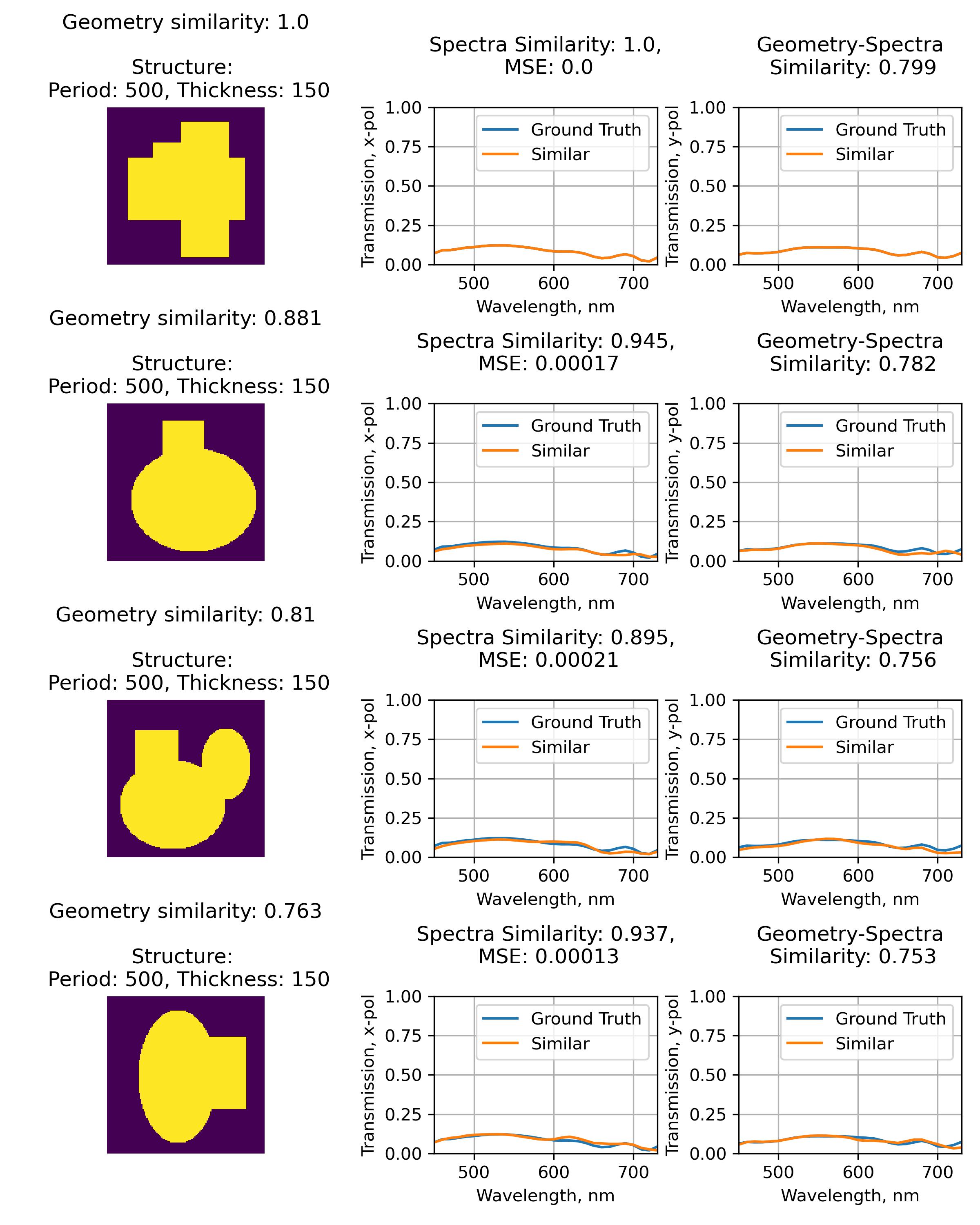}
	    \caption{\textbf{Non-uniqueness analysis for a complex geometry.} }
     \label{fig:nonuniq2}
\end{figure}

\begin{figure}[!htbp] \centering
\includegraphics[width=\linewidth,height=0.78\textheight,keepaspectratio]{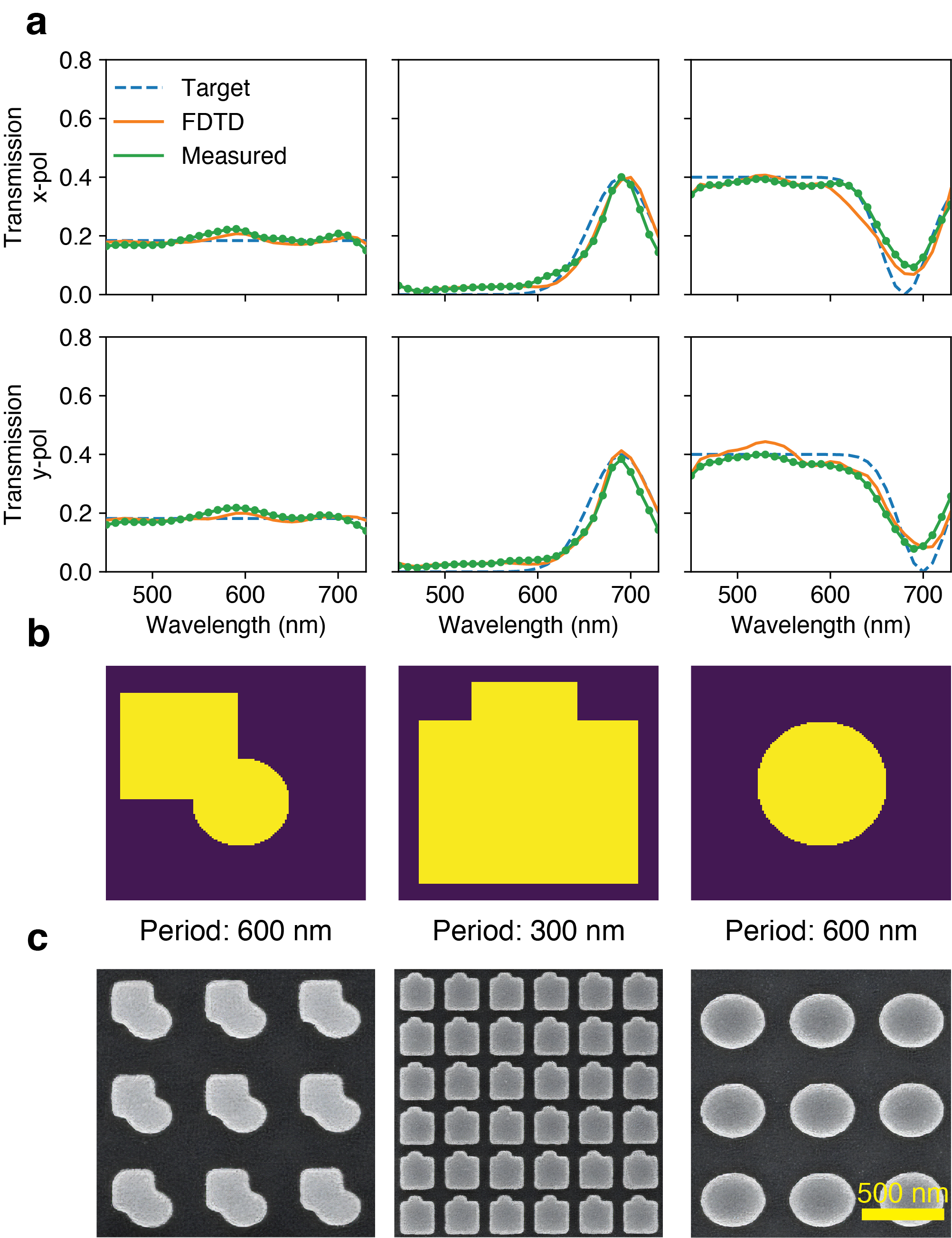}
	    \caption{\textbf{Experimental validation of design by latent space optimization.}
        \textbf{a}.~Comparison of transmission spectra for x- and y-polarizations, showing the inverse design targets, FDTD simulated spectra, and measured spectra. \textbf{b}.~Predicted designs with a thickness of \SI{150}{\nano\meter}. \textbf{c}.~SEM images of the fabricated structures.}
     \label{fig:optim}
\end{figure}

\begin{figure}[!htbp] \centering
\includegraphics[width=\linewidth,height=0.78\textheight,keepaspectratio]{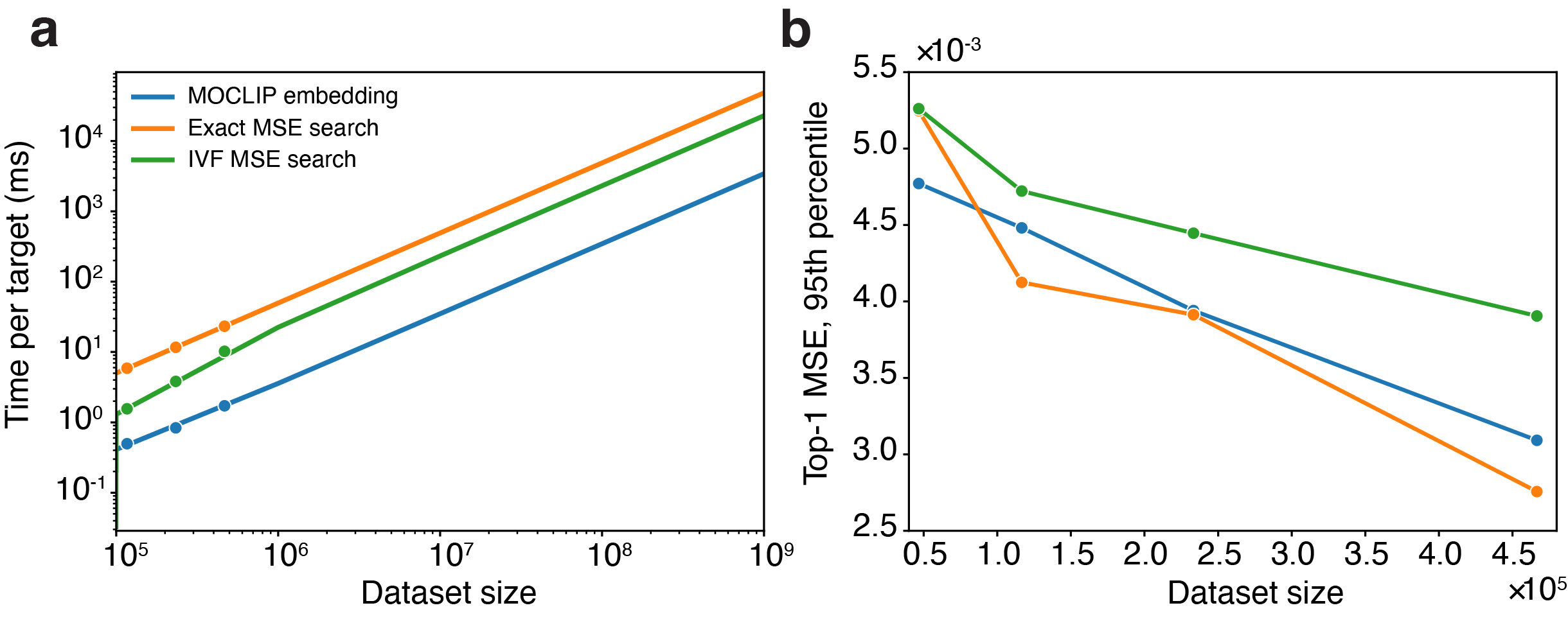}
	    \caption{\textbf{Library search scaling trend.}
\textbf{a}.~Comparison of time per target as a function of dataset size for MOCLIP embedding, exact MSE search, and IVF MSE search. \textbf{b}.~Top-1 accuracy comparison across the three approaches.}
     \label{fig:library_search}
\end{figure}